\documentclass{aa}
\usepackage{graphicx}
\usepackage{natbib}

\bibpunct[]{(}{)}{;}{a}{}{,}
\def\as{\mbox{$a_{\mbox{\tiny s}}$}}
\def\rout{\mbox{$R_{\mbox{\tiny out}}$}}
\def\ltsima{$\; \buildrel < \over \sim \;$}
\def\simlt{\lower.5ex\hbox{\ltsima}}
\def\gtsima{$\; \buildrel > \over \sim \;$}
\def\simgt{\lower.5ex\hbox{\gtsima}}

\def\macc{\mbox{$\dot{M}_{\mbox{\Large acc}}$}}

\def\macc{$\dot{M}_{\rm acc}$~}
\def\maccm{\dot{M}_{\rm acc}}

\def\cm2{\mbox{$\mbox{cm}^{-2}$}}
\def\cm3{\mbox{$\mbox{cm}^{-3}$}}

\begin{document}
   \title{Molecular line study of the very young protostar\\
          IRAM 04191 in Taurus: Infall, rotation, and outflow}

   \author{A. Belloche
          \inst{1}
          \and
          P. Andr\'e\inst{1}
          \and
	  D. Despois\inst{2}
          \and
          S. Blinder\inst{2,3}
          }

   \offprints{belloche@discovery.saclay.cea.fr or pandre@cea.fr}

   \institute{Service d'Astrophysique, CEA/DSM/DAPNIA, C.E. Saclay, 
              F-91191, Gif-sur-Yvette Cedex, France
         \and
              Observatoire de Bordeaux (INSU/CNRS),  B.P. 89,
              F-33270 Floirac, France
         \and
              Division of Nuclear Medicine, 
              Vancouver Hospital and Health Sciences Center,
              Vancouver, B.C., Canada
             }

   \date{Received April 16, 2002; accepted July 12, 2002}

   \abstract{We present a detailed millimeter line study of the circumstellar 
environment of the low-luminosity Class 0 protostar IRAM 04191~+~1522 in the 
Taurus molecular cloud. 
New line observations demonstrate that the $\sim 14000$~AU radius
protostellar envelope is 
undergoing both extended infall and fast, differential rotation. Radiative 
transfer modeling of multitransition CS and C$^{34}$S maps  
indicate an infall velocity $v_{\rm inf} \sim 0.15$ km~s$^{-1}$ at 
$r \sim 1500$ AU and $v_{\rm inf} \sim 0.1$ km~s$^{-1}$ up to $r \sim 11000$~AU,  as well as 
a rotational angular velocity $\Omega \sim 3.9 \times 10^{-13}$~ rad~s$^{-1}$,
strongly decreasing with radius beyond 3500~AU down to a value 
$\Omega \sim $~1.5--3$\times 10^{-14}$~rad~s$^{-1}$ at $\sim $~11000~AU.
Two distinct regions, which differ in both their infall and their 
rotation properties, therefore seem to stand out: 
the inner part of the envelope ($r \simlt 2000-4000$ AU) is 
rapidly collapsing and rotating, while the outer part undergoes only 
moderate infall/contraction and slower rotation. 
These contrasted features suggest that angular momentum is conserved 
in the collapsing inner region but efficiently dissipated due to  
magnetic braking in the slowly contracting outer region.
We propose that the inner 
envelope is in the process of decoupling from the ambient cloud and 
corresponds to the effective mass reservoir ($\sim 0.5\, M_\odot $) 
from which the central star is being built. 
Comparison with the rotational properties of other 
objects in Taurus suggests that IRAM~04191 is at a pivotal 
stage between a prestellar regime of constant angular velocity enforced by 
magnetic braking and a dynamical, protostellar regime of nearly conserved 
angular momentum. The rotation velocity profile we derive for the inner 
IRAM~04191 envelope should thus set some constraints on the distribution 
of angular momentum on the scale of the outer Solar system at the onset 
of protostar/disk formation.
    }
   \titlerunning{Velocity structure of the IRAM 04191 envelope}
   \authorrunning{Belloche et al.}
   \maketitle
%

\section{Introduction}
\label{intro}

\subsection{The enigmatic onset of protostellar collapse}
\label{intro_background}

Despite recent progress, the initial conditions of star formation and 
the first phases of protostellar collapse remain poorly known 
\citep[e.g.][ for reviews]{Myers99,Andre00}.
In the standard theory of isolated, low-mass star 
formation \citep[e.g.][]{Shu87}, the initial conditions 
correspond to essentially static singular isothermal spheres 
(SISs, which have $\rho = (\as^2/2\pi G )r^{-2}$), assumed to be in slow, 
solid-body rotation \citep[][ -- TSC84]{Terebey84}.
This leads to a strictly constant mass accretion rate, $\maccm \sim \as^3/G$
(where $\as$ is the isothermal sound speed), and to a growth of the 
centrifugal disk as $R_{disk} \propto  t^3$ (cf. \citeauthor*{Terebey84}) 
during the protostellar accretion phase ($t > 0$).
Other theoretical models exist, however, that predict 
a time-dependent accretion history 
if the collapse initial conditions are either not singular or not scale-free 
\citep[e.g.][]{Foster93,Henriksen97,Basu97,Ciolek98,Hennebelle02}.
Starting from realistic, finite-sized prestellar cores 
with $\rho \approx cte$ in their central region 
\citep[c.f.][]{Ward99,Bacmann00,Alves01}, these models yield supersonic 
inward velocities close to the center prior to point mass formation 
(i.e. at $t<0$) and result in denser, nonequilibrium density
distributions with strong differential rotation 
at the onset of the main accretion phase, i.e., at $t=0$.   
In these models, the accretion rate \macc is initially significantly larger  
than in the Shu model, then 
quickly converges toward the standard $\sim \as^3/G$ value, and finally  
declines much below $\as^3/G$ because of the finite  
reservoir of mass \citep*[see, e.g.,][]{Foster93,Henriksen97}. 
Conservation of angular momentum during dynamical collapse at $t<0$ produces 
a differential rotation profile at $t=0$ \citep[e.g. $\Omega \propto r^{-1}$ 
in the magnetically-controlled model of][]{Basu98}. This rotation profile in 
turn implies a more rapid growth of $R_{disk}$ initially 
(i.e., $R_{disk} \propto  t$ at small $t>0$ in the Basu model) than 
in the \citeauthor*{Terebey84} model.

Getting at a better, more quantitative knowledge of protostellar collapse 
is crucial, e.g., to gain insight into the origin of stellar masses and 
disk formation. Observationally, there are two complementary approaches to
estimating the initial conditions of protostar formation. 
The first approach consists in studying the structure and kinematics 
of ``prestellar cores'' such as L1544 \citep*[e.g.][]{Ward99,Tafalla98}
, representative of times $t \simlt 0$. The second approach, adopted here, is 
the detailed study of Class~0 accreting protostars observed at $t \simgt 0$, 
such as IRAM~04191 (see \S~\ref{intro_iram04191}),  
which should still retain detailed memory of their initial conditions.

\subsection{IRAM~04191: A  very young Class 0 protostar in Taurus}
\label{intro_iram04191}

\begin{figure} [!ht]
\resizebox{\hsize}{!}{\includegraphics[width=3.5cm,angle=0]{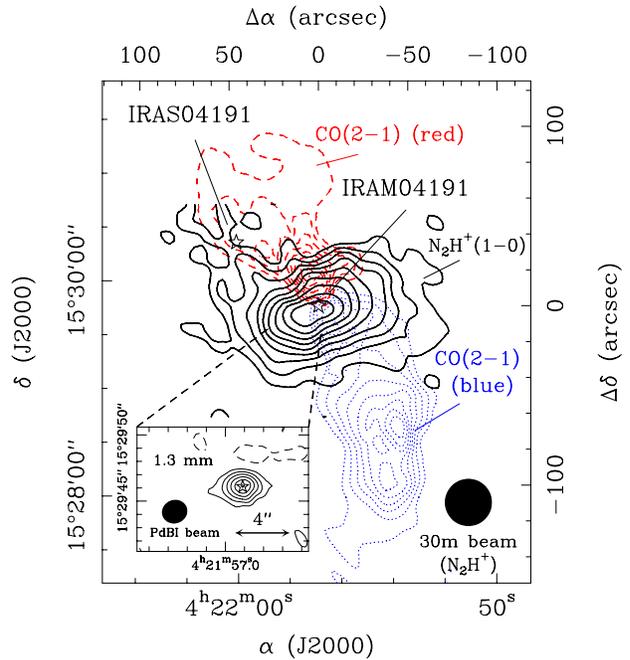}}
 \caption[]{N$_2$H$^+$(1-0) integrated intensity map of the IRAM~04191
protostellar envelope overlaid on the CO(2-1) outflow map of 
\citeauthor*{Andre99}.
Both maps were taken with the IRAM 30m telescope. The 
N$_2$H$^+$(1-0) emission is integrated over the whole seven-component 
multiplet from 5.7 to 22.6 km~s$^{-1}$ and the contours go from 0.5 to 4.5 
K~km~s$^{-1}$ by 0.5  K~km~s$^{-1}$. The blueshifted CO(2-1) emission 
({\it dotted contours}) is integrated over the 0-5 km~s$^{-1}$ velocity 
range, and the redshifted CO(2-1) emission ({\it dashed contours}) is 
integrated over the 8-13 km~s$^{-1}$ velocity range; the CO(2-1) contours go 
from 5 to 21 K~km~s$^{-1}$ by 2 K~km~s$^{-1}$.\\ 
The insert in the bottom-left corner 
shows a 227~GHz dust continuum map of the inner envelope obtained
at the IRAM Plateau de Bure interferometer (see \S~\ref{obs_pdbi}), with 
positive (solid) contours from $+1$ to $+6$ by 1 mJy/1.9\arcsec -beam; 
the dotted contour is negative at $-1$~mJy/1.9\arcsec -beam). \\
In both maps, the central star symbol at (0,0) marks the position 
of IRAM~04191 as determined by a 2D Gaussian fit to the PdBI 227~GHz 
continuum image. The other star symbol indicates the position of the 
Class I source IRAS~04191.
}
 \label{n2h+flow}
\end{figure}

The massive ($M_{tot} \sim 1.5\, M_\odot $) dense core/envelope of the
Class~0 object, IRAM 04191~+~1522 (hereafter IRAM 04191), was 
originally discovered in the millimeter dust continuum with the IRAM 30m
telescope in the southern part of the 
Taurus molecular cloud \citep[][ -- hereafter AMB99]{Andre99}. 
Follow-up observations revealed a highly 
collimated CO bipolar outflow (see Fig.~\ref{n2h+flow}), a weak 3.6~cm VLA 
radio continuum source located at its center of symmetry, 
and spectroscopic evidence of spatially extended infall motions in the bulk of
the envelope.
These are typical attributes of a Class~0 
protostar \citep[][]{Andre93,Andre00}.

The very high envelope mass to luminosity ratio of IRAM 04191 
($M_\mathrm{env}^{< 4200 \mathrm{ AU}}/L_\mathrm{bol}\simgt 3$  
M$_\odot$/L$_\odot$) and its position in the 
$L_\mathrm{bol}-T_\mathrm{bol}$ evolutionary diagram 
($L_\mathrm{bol} \sim 0.15$ L$_\odot$ ~and $T_\mathrm{bol} \sim 18$ K) 
suggest an age $t \sim 1-3 \times 10^4$ yr since the beginning of the 
accretion phase (see \citeauthor*{Andre99}). This is significantly younger 
than all of the IRAS candidate protostars of Taurus \citep[e.g.][]{Kenyon93}, 
including L1527 which has  $t \simlt 10^5$~yr \citep[e.g.][]{Ohashi97a}.
IRAM~04191 thus appears to be the youngest accreting protostar known so far 
in Taurus, although the collapsing
protostellar condensation MC~27 discovered by \citet{Onishi99}
may be in a comparable evolutionary state.

As IRAM~04191 is particularly young, nearby (d $= 140$ pc), and 
relatively isolated, the study of its velocity structure 
based on molecular line observations provides a unique opportunity to set 
constraints on collapse models.
This is especially true since the viewing angle is favorable. The CO(2--1) 
outflow map of \citeauthor*{Andre99} (see Fig.~\ref{n2h+flow}) shows 
well separated outflow lobes with almost no overlap between blue-shifted and 
red-shifted emission, indicating that the flow lies out of the plane of the
sky at an intermediate inclination angle \citep[e.g.][]{Cabrit90}.
In addition, \citeauthor*{Andre99} estimate an aspect ratio of 
$\sim 0.65$ for the circumstellar dust/N$_2$H$^+$ envelope, whose major axis
is perpendicular to the outflow axis (cf. Fig.~\ref{n2h+flow}).
Both characteristics are consistent with an inclination angle 
of the outflow axis to the line of sight of $i \sim 50 \degr$.

Here, we present and discuss the results of a comprehensive set of molecular
line observations toward IRAM~04191. 
The layout of the paper is as follows. Sect.~\ref{obs_set} summarizes  
observational details. Sect.~\ref{obs_ana} interprets the observations in 
terms of infall, rotation, and outflow motions in the envelope. 
We then model the observed spectra using   
radiative transfer simulations computed in 1D spherical geometry 
with radial infall motions (Sect.~\ref{simul_1d}) and 
in 2D axial geometry with both infall and rotational motions
(Sect.~\ref{simul_2d}).  
Sect.~\ref{discuss} compares the derived constraints on the velocity
structure of the IRAM~04191 envelope with the predictions of collapse models. 
Our conclusions are summarized in Sect.~\ref{concl}.

\section{Observations}
\label{obs_set}

We carried out  millimeter line observations with the IRAM 30m 
telescope at Pico Veleta, Spain, during 7 nights in July and August 1999, in 
the following molecular transitions: N$_2$H$^+$(1-0), CS(2-1), 
C$^{34}$S(2-1), HCO$^+$(1-0), H$^{13}$CO$^+$(1-0), 
C$_3$H$_2$(2$_{12}$-1$_{01}$) at 3~mm, CS(3-2), C$^{34}$S(3-2), 
H$_2$CO(2$_{12}$-1$_{11}$) at 2~mm, and CS(5-4), H$_2$CO(3$_{12}$-2$_{11}$), 
C$^{18}$O(2-1), DCO$^+$(3-2) at 1.2~mm. 
Our adopted set of line frequencies is given in Table \ref{tab_freq}. 
The half-power beamwidths were $\sim 26 ~\arcsec$, $\sim 17 ~\arcsec$ and 
$\sim 10 ~\arcsec$ at 3~mm, 2~mm and 1.2~mm, respectively. 
We used four SIS heterodyne receivers simultaneously and an autocorrelation
spectrometer as backend, with a spectral resolution of 20~kHz at 3~mm and 
40~kHz at 2~mm and 1.2~mm. The corresponding velocity resolutions ranged from 
0.05 to 0.08 km~s$^{-1}$ per channel. 
The observations were done in single sideband mode, with sideband rejections 
of 0.01, 0.1 and 0.05 at 3~mm, 2~mm and 1.2~mm, respectively.
The forward and beam efficiencies of the telescope 
used to convert antenna temperatures $T^*_\mathrm{A}$ into 
main beam temperatures $T_\mathrm{mb}$ are listed in Table~\ref{tab_freq}. 
The system temperatures 
($T^*_\mathrm{A}$ scale) ranged from $\sim 110$ K to $\sim 150$ K at 3~mm, 
$\sim 280$ K to $\sim 410$ K at 2~mm, and $\sim 300$ K to $\sim 550$ K at 
1.2~mm. The telescope pointing was checked every $\sim 2$ hours 
on Saturn, 0528+134, and/or 0420-014  and found to be accurate to 
$\sim 3 ~\arcsec$ (rms).
The telescope 
focus was optimized on Saturn every $\sim 3$ hours. Position switching 
observations were done with a reference position located at either
($\Delta\alpha$,$\Delta\delta$)~=~($1200 ~\arcsec$,$-1200 ~\arcsec$) 
or ($80 ~\arcsec$,$-80 ~\arcsec$) relative to the (0,0) position 
(envelope center as measured in the 1.3mm continuum).
Extensive mapping was performed in the ``on-the-fly'' mode 
\citep[][]{Ungerechts00}. A few additional C$^{18}$O(1-0), C$^{18}$O(2-1), 
CS(3-2), C$^{34}$S(3-2), CS(5-4) and H$^{13}$CO$^+$(3-2) position-switch 
observations, performed in September 1997, November 2000
and October 2001, will also be used here.
All of these single-dish data were reduced with the CLASS software package 
\citep[][]{Buisson02}.

In addition, we also observed IRAM~04191 with the 5-antenna IRAM Plateau 
de Bure interferometer (PdBI)
in its B1, C2, D configurations between January and April 1999.
The two receivers of each antenna were tuned to the CS(2-1) and 
H$_2$CO(3$_{12}$-2$_{11}$) lines, with spectral resolutions of 40 kHz and 80 
kHz, corresponding to velocity resolutions of 0.12 km~s$^{-1}$ and 0.10
km~s$^{-1}$, respectively. 
The four remaining windows of the PdBI correlator were used 
to record the continuum emission with a total bandwidth of 300~MHz 
at both 98~GHz ($\lambda \sim 3$~mm) and 
227~GHz ($\lambda \sim 1.3$~mm). 
The (naturally-weighted) synthesized half-power beamwidths were 
$4.5~\arcsec \times 4.4~\arcsec$ (630 AU $\times$ 620 AU) at 98 GHz and 
$1.9~\arcsec \times 1.8~\arcsec$ (270 AU $\times$ 250 AU) at 227 GHz, 
and the (FWHP) primary beams $\sim 50 ~\arcsec$ and $\sim 25 ~\arcsec$, 
respectively. The 
correlator bandpass was calibrated on the strong source 3C273. 
Several nearby phase calibrators were observed to determine the time-dependent
complex antenna gains. The absolute calibration uncertainty was estimated to 
be $\sim 15\% $. The data were calibrated and imaged using the CLIC 
\citep[][]{Lucas99} and Mapping \citep[][]{Guilloteau02} packages in the IRAM 
software.

\begin{table}
\centering
 \caption[]{Adopted line rest frequencies and telescope efficiencies.}
 \label{tab_freq}
 \begin{tabular}{|l|rcccc|}
 \hline
 \multicolumn{1}{|c|}{Line} & \multicolumn{1}{c}{Frequency$^{(1)}$ } & \hspace*{-0.5cm} $\sigma_\mathrm{v}$$^{(2)}$ & \hspace*{-0.5cm} Ref. & \hspace*{-0.3cm} F$_\mathrm{eff}$$^{(4)}$ & \hspace*{-0.4cm} B$_\mathrm{eff}$$^{(4)}$ \\
  & \multicolumn{1}{c}{(MHz)} & \hspace*{-0.5cm} {\scriptsize (km~s$^{-1}$)} & \hspace*{-0.5cm} $^{(3)}$ & \hspace*{-0.3cm} ($\%$) & \hspace*{-0.4cm} ($\%$) \\
 \hline\hline
C$_3$H$_2$(2$_{12}$-1$_{01}$)      & $85338.905(6)$    & \hspace*{-0.5cm} 0.02 & \hspace*{-0.5cm} (1) & \hspace*{-0.3cm} 92 & \hspace*{-0.4cm} 73 \\
N$_2$H$^+${\scriptsize (101-012)}  & $93176.258(7)$   & \hspace*{-0.5cm} 0.02 & \hspace*{-0.5cm} (2) & \hspace*{-0.3cm} 92 & \hspace*{-0.5cm} 73 \\
C$^{34}$S(2-1)                     & $96412.952(1)$    & \hspace*{-0.5cm} 0.003 & \hspace*{-0.5cm} (3)  & \hspace*{-0.3cm} 92 & \hspace*{-0.4cm} 73 \\
CS(2-1)                            & $97980.953(1)$    & \hspace*{-0.5cm} 0.003 & \hspace*{-0.5cm} (3)  & \hspace*{-0.3cm} 90 & \hspace*{-0.4cm} 73 \\
C$^{18}$O(1-0)                     & $109782.175(2)$   & \hspace*{-0.5cm} 0.005 & \hspace*{-0.5cm} (4)  & \hspace*{-0.3cm} 92 & \hspace*{-0.4cm} 73 \\
C$^{34}$S(3-2)                     & $144617.101(1)$   & \hspace*{-0.5cm} 0.002 & \hspace*{-0.5cm} (3)  & \hspace*{-0.3cm} 87 & \hspace*{-0.4cm} 65 \\
                                   &                   & \hspace*{-0.5cm}   & \hspace*{-0.5cm}      & \hspace*{-0.3cm} 90 & \hspace*{-0.4cm} 54 \\
CS(3-2)                            & $146969.026(1)$   & \hspace*{-0.5cm} 0.002 & \hspace*{-0.5cm} (3)  & \hspace*{-0.3cm} 90 & \hspace*{-0.4cm} 54 \\
C$^{18}$O(2-1)                     & $219560.3541(15)$ & \hspace*{-0.5cm} 0.002 & \hspace*{-0.5cm} (5)  & \hspace*{-0.3cm} 86 & \hspace*{-0.4cm} 42 \\
CS(5-4)                            & $244935.555(1)$   & \hspace*{-0.5cm} 0.001 & \hspace*{-0.5cm} (3)  & \hspace*{-0.3cm} 84 & \hspace*{-0.4cm} 42 \\
 \hline
 \end{tabular}
 \begin{list}{}{}
 \item[Notes: $^{(1)}$]{The frequency uncertainty in units of the last
significant digit is given in parentheses.}
 \item[$^{(2)}$]{Frequency uncertainty converted in units of velocity.}
 \item[$^{(3)}$ References for rest frequencies:]{(1) Laboratory measurement
   from 
   \citet{Vrtilek87}; (2) Observational result from \citet{Lee01}; (3) 
   Laboratory measurement from \citet{Gottlieb02}. (4) Laboratory measurement 
   from Klapper (2001, private communication). (5) Laboratory measurement from 
   \citet{Klapper01}.}
 \item[$^{(4)}$]{Forward and beam efficiencies of the IRAM 30m telescope
   (\citet{Wild99} and http://www.iram.es/).}
 \end{list}
\end{table}

\section{Results and qualitative interpretation : Evidence for rotation and 
infall}
\label{obs_ana}

%
\subsection{Weak 1.3~mm continuum detection at PdBI}
\label{obs_pdbi}

The PdBI continuum image at 227~GHz (see Fig.~\ref{n2h+flow}) reveals weak, 
point-like emission centered at 
$\alpha_\mathrm{(2000)}= 04^\mathrm{h}21^\mathrm{m}56^\mathrm{s}_{^.}91$, 
$\delta_\mathrm{(2000)}= 15\degr 29\arcmin 46.1\arcsec$, a position which
should be accurate to better than $0.5\arcsec$. (The PdBI position is offset 
by $\sim 5 ~\arcsec$ from the N$_2$H$^+$(1-0) centroid observed at the 30m 
telescope, which is only marginally significant given the $\sim 3 ~\arcsec$ 
single-dish pointing accuracy.) We measure a peak 227~GHz flux density of 
$S_\mathrm{peak}^{1.9\arcsec} = 6.1 \pm 0.4$ mJy/$1.9\arcsec$-beam
at the central position. 
The 227~GHz emission is slightly resolved and a Gaussian fit performed in the
uv-plane 
yields a deconvolved FWHM size of 
$(1.1 \arcsec \pm 0.4 \arcsec) \times (0.6 \arcsec \pm 0.3 \arcsec)$ with a
position angle P.A. $= 84~\degr$. At 98 GHz, the rms noise level is 
0.14 mJy/$4.5\arcsec$-beam and we do not detect any emission above 
0.6 mJy/$4.5\arcsec$-beam ($\sim 4\, \sigma $)\footnote{We measure a peak
98~GHz flux density of 0.6 mJy/$4.5\arcsec$-beam close to ($\sim 4\arcsec$ 
north-east of) the central 227~GHz position but the cleaned image contains 
negative contours down to $-0.8$~mJy/$4.5\arcsec$-beam.}. 

For comparison, using the IRAM 30m telescope equipped with the MPIfR bolometer
array (MAMBO), \citet[][]{Motte01} measured a peak flux density
of $S_\mathrm{peak}^{11\arcsec} = 110 \pm 7$ mJy/$11\arcsec$-beam 
at 1.3~mm \citep[$\nu_{eff} \sim 240$~GHz -- e.g.][]{Broguiere02} 
and a radial intensity profile of the form 
$I(\theta) \propto \theta^{-0.6 \pm 0.1}$ 
in the range of angular radii $\theta = 11~\arcsec$ to $\theta = 100~\arcsec$. 
This extended 1.3~mm continuum source is clearly 
the dust counterpart of the circumstellar gas envelope/core observed in 
N$_2$H$^+$ (cf. Fig.~\ref{n2h+flow}).
Assuming the same radial intensity profile holds at smaller angular radii,
one expects the peak flux density of the envelope to scale as 
$S_\mathrm{peak}(\theta_b) \propto \theta_b^{1.4 \pm 0.1}$ with beamsize 
$\theta_b $. If we adopt a dust opacity index $\beta = 1.5$ to account for
the slight difference in observing frequency between the 30m and PdBI
measurements, this flux-density scaling predicts
$S_\mathrm{peak}^{1.9\arcsec} \approx 7.7$~mJy/$1.9\arcsec$-beam, which is
only 25\%  larger than the PdBI peak flux density quoted above. 
Given the relative calibration uncertainties, this comparison suggests that
the weak 227~GHz emission detected at PdBI arises from the inner part of the
envelope seen at the 30m telescope rather than from an  
accretion disk surrounding the central protostellar object.
Furthermore, a single power law $I(\theta) \propto \theta^{-0.5 \pm 0.2}$ 
appears to characterize the radial intensity profile of the envelope
over the whole range of angular radii from $\sim 1 \arcsec$ to 
$\sim 100 \arcsec$. Such an intensity profile 
implies either a relatively flat $\rho \propto r^{-1}$ density profile
if central heating with 
$T_\mathrm{dust} \propto r^{-0.4}$ applies in the inner part of the envelope
(see Fig.~\ref{model_1d} below), or a $\rho \propto r^{-1.5}$ density profile 
if the dust temperature is uniform.\\
The weak emission detected at PdBI sets strong constraints on the mass and 
size of any central accretion disk. 
Assuming optically thin emission, a dust mass opacity of 0.02 cm$^2$~g$^{-1}$ 
typical of circumstellar disks \citep[cf.][]{Beckwith90} and a mean dust 
temperature of 20 K, the PdBI peak flux density yields 
$M_\mathrm{disk} < 1 \times 10^{-3}$ M$_\odot$. 
Alternatively, the assumption of optically thick dust emission at 20~K 
implies $R_\mathrm{disk} < 10$ AU. 

%
\subsection{Fast, differential rotation}
\label{obs_rotation}

\begin{figure}
\resizebox{\hsize}{!}{\includegraphics[angle=270]{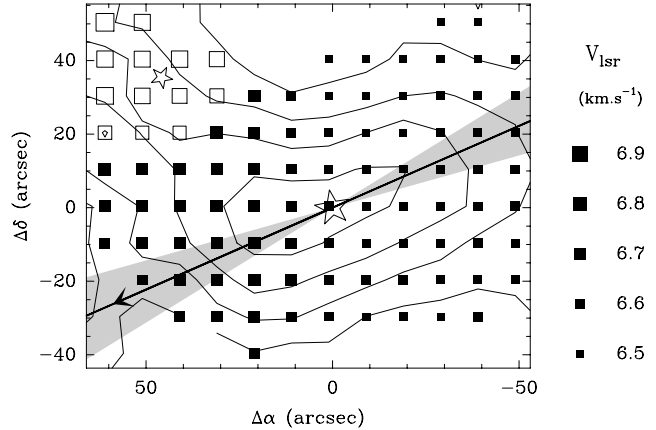}}
 \caption{Map of the peak LSR velocity ({\it filled squares}) in the 
IRAM~04191 envelope as derived from Gaussian fits to the 
C$_3$H$_2$(2$_{12}$-1$_{01}$) spectra observed with the 30m telescope. 
The map of integrated C$_3$H$_2$(2$_{12}$-1$_{01}$) intensity 
between 5.7 and 7.7 km~s$^{-1}$ is overlaid as contours 
(ranging from 0.4 to 1.0 by 0.2 K~km~s$^{-1}$).
The black arrow shows the best-fit direction 
(P.A. $\sim 114\degr \pm 8\degr$) of the mean velocity gradient 
($\sim 3.4$ km~s$^{-1}$~pc$^{-1}$) across the envelope; the grey-shaded 
sectors display the 1$\sigma$ error on this orientation.
The size of each square is proportional to the LSR velocity at its position.
The open squares in the upper-left corner give the LSR velocities measured 
at positions that were ignored in fitting the velocity gradient because of  
likely contamination by the Class~I source IRAS~04191 and/or the 
redshifted lobe of the IRAM~04191 outflow.
The two star symbols mark the positions of IRAM~04191 [at (0,0)]
and IRAS~04191.}
 \label{map_vcent}
\end{figure}

\begin{figure}
\resizebox{\hsize}{!}{\includegraphics[angle=270]{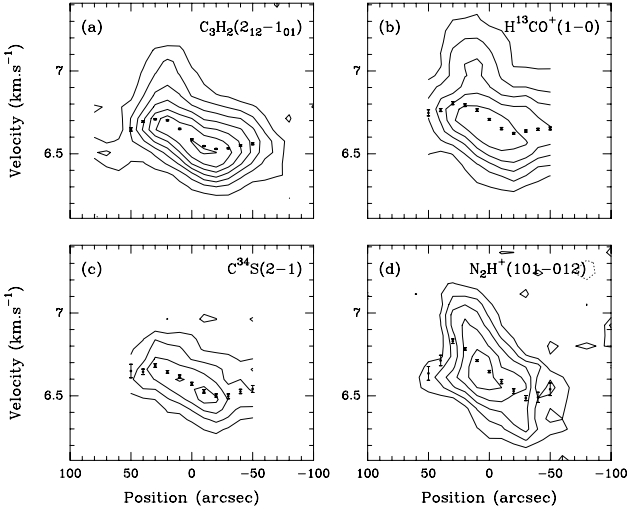}}
\caption{Position-velocity diagrams in small optical depth lines along the 
direction perpendicular to the outflow axis. 
(a) C$_3$H$_2$(2$_{12}$-1$_{01}$); (b) H$^{13}$CO$^+$(1-0); 
(c) C$^{34}$S(2-1); (d) N$_2$H$^+$(101-012). 
The dots with error bars represent the peak velocities as determined by 
Gaussian fits. The base contour and contour step have the same value:
0.3~K (in T$_\mathrm{A}^*$) in (a) and (b); 0.2 K in (c) and~(d). 
Note how the dots mark a similar S shape in all four diagrams, suggestive
of differential rotation in the envelope (see text).}
\label{pvdiag}
\end{figure}

All of the centroid velocity (first-order moment) maps taken 
at the 30m telescope in small optical depth lines, such 
as C$_3$H$_2$(2$_{12}$-1$_{01}$), N$_2$H$^+$(101-012), H$^{13}$CO$^+$(1-0),  
and C$^{34}$S(2-1), show a clear velocity 
gradient across the envelope (see, e.g., Fig.~\ref{map_vcent}).  
The (south)-east part is redshifted with respect to the source systemic 
velocity, while the (north)-west part is blueshifted. 
We have applied the sector method described by 
\citet{Arquilla86} to measure the direction of this velocity gradient in the 
C$_3$H$_2$(2$_{12}$-1$_{01}$) map using Gaussian fits to the spectra. We 
obtain a position angle P.A. $\sim 114\degr \pm 8\degr$ 
\footnote{Before fitting the direction of the velocity gradient, 
we have masked the redshifted emission seen in the north-east part 
of the C$_3$H$_2$(2$_{12}$-1$_{01}$) map shown in Fig.~\ref{map_vcent}, 
probably due to the redshifted lobe of the outflow or to the nearby Class I 
source IRAS~04191+1523 \citep{Tamura91}.}
(cf. Fig.~\ref{map_vcent}).
The velocity 
gradient thus lies along the major axis of the elongated dust/N$_2$H$^+$ 
core (P.A. $\sim 120\degr$), i.e., perpendicular to the outflow axis. 
Coupled to the high degree of symmetry of the position-velocity 
diagrams shown in Fig.~\ref{pvdiag} with respect to envelope center,
this strongly suggests that the envelope is rotating 
about an axis coinciding with the outflow axis (cf. Fig.~\ref{n2h+flow}).
Turbulent motions would produce a more random velocity field 
\citep[cf.][]{Burkert00} and are weak here anyway (see \S~\ref{obs_width} and
Fig.~\ref{model_vel}b below).

From now on, we adopt a position angle P.A.~$= 28\degr$ 
for the projection of the rotation/outflow axis onto the plane of the sky. 
To minimize contamination by the outflow, we analyze 
the velocity structure of the envelope along the axis perpendicular to the 
outflow (P.A.~$= 28\degr$) and going through the center. 
Along this axis, the magnitude of the velocity 
gradient estimated from linear fits to the C$_3$H$_2$(2$_{12}$-1$_{01}$),
H$^{13}$CO$^+$(1-0), C$^{34}$S(2-1), and N$_2$H$^+$(101-012) 
position-velocity diagrams is $\sim 6$~km~s$^{-1}$~pc$^{-1}$, $\sim
7$~km~s$^{-1}$~pc$^{-1}$, $\sim 6$~km~s$^{-1}$~pc$^{-1}$, 
and 10~km~s$^{-1}$~pc$^{-1}$, respectively, increasing from north-west 
to south-east over $40 ~\arcsec$ (see Fig.~\ref{pvdiag}). 
The mean velocity gradient is $7 \pm 2$ km~s$^{-1}$~pc$^{-1}$. 
Given the estimated viewing angle $i = 50\degr$ of the flattened envelope 
(cf. \S~\ref{intro_iram04191}), this implies a mean rotational angular 
velocity of $\Omega = 9 \pm 3$ km~s$^{-1}$~pc$^{-1}$
in the inner $r \sim 2800$~AU radius region\footnote{With such an angular 
velocity, the gas at $r \sim 2800$~AU would make a complete turn in 
$\sim 0.7$~Myr, which is comparable to the typical lifetime of prestellar 
cores with central densities $\simgt 10^5$~cm$^{-3}$ 
\citep[e.g.][]{Jessop00}. Thus, although the observed velocity 
gradient is probably not indicative of a well-developed circular motion, 
it may have already induced significant rotational distortion in the envelope.
}.

The C$_3$H$_2$(2$_{12}$-1$_{01}$), H$^{13}$CO$^+$(1-0), and 
N$_2$H$^+$(101-012) position-velocity diagrams shown in Fig.~\ref{pvdiag} 
may be slightly contaminated by a secondary, redshifted component toward 
the east. We have used the GAUSSCLUMPS algorithm of \citet{Stutzki90}
\citep[see also][]{Kramer98} to try and subtract
this secondary component. The algorithm finds a Gaussian component at 
approximately ($+50~\arcsec$, $+30~\arcsec$) in 
C$_3$H$_2$(2$_{12}$-1$_{01}$) and ($+30\arcsec$, $+20\arcsec$) in 
H$^{13}$CO$^+$(1-0). 
The removal of this secondary component reduces the magnitude of the 
velocity gradient measured in C$_3$H$_2$(2$_{12}$-1$_{01}$) 
and H$^{13}$CO$^+$(1-0) by $\sim 30\%$. 
GAUSSCLUMPS fails to identify any secondary component near the 
same location in the N$_2$H$^+$(101-012) data. 
The velocity gradient thus seems to be significantly larger in 
N$_2$H$^+$(101-012) than in the other three lines (see below).

\begin{figure}
\resizebox{\hsize}{!}{\includegraphics[angle=270]{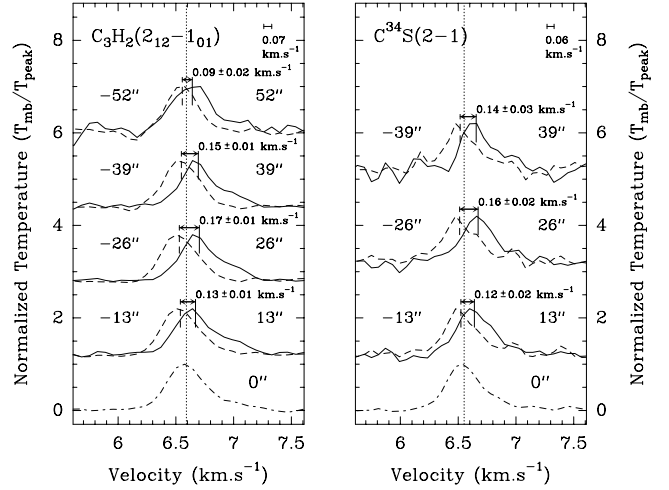}}
\caption{Normalized C$_3$H$_2$(2$_{12}$-1$_{01}$) and C$^{34}$S(2-1) spectra
observed every half beamwidth along the direction perpendicular to 
the outflow axis. Each spectrum has been divided by its peak temperature 
for easier inter-comparison.  
In both panels, the vertical dotted line marks the peak velocity of the 
central spectrum ({\it dash-dotted curve} at the bottom) measured by a 
Gaussian fit. The double-sided arrows and associated values indicate the 
magnitude of the velocity shifts (determined by Gaussian fits) between 
spectra observed at symmetric positions ($\pm 13\arcsec$, $\pm 26 \arcsec$,
$\pm 39 \arcsec$, or $\pm 52 \arcsec$) about the center. 
The channel resolution is shown in the 
upper right corner of each panel.
While solid-body rotation would produce a velocity shift increasing 
linearly with radius, {\it the opposite trend is observed} ~beyond one 
beamwidth ($26 \arcsec$) from the center: the velocity shift
decreases for larger offsets about the central position. This strongly
suggests the presence of {\it differential rotation} in the envelope.
}
\label{spec_rot}
\end{figure}

Remarkably, all of the position-velocity diagrams shown here present a S 
shape, clearly seen on the peak velocity curves derived from Gaussian fits 
(see Fig.~\ref{pvdiag}). 
On either side of the source, the velocity shift with respect to the 
systemic velocity increases in absolute value up to an angular 
radius $\theta_\mathrm{m} \sim 25 \pm 10\arcsec$ 
(i.e., one full beamwidth of the 30m telescope at 3mm, corresponding to 
$r_\mathrm{m} \sim 3500 \pm 1400$ AU), and then decreases for 
$\theta > \theta_\mathrm{m}$. This is further illustrated in 
Fig.~\ref{spec_rot} which shows the C$_3$H$_2$(2$_{12}$-1$_{01}$) and 
C$^{34}$S(2-1) spectra observed every half beamwidth (Nyquist sampling)
at symmetric positions with respect to 
source center along the direction perpendicular to the outflow axis.
It can be seen that the velocity shift between symmetric positions decreases
from $0.17$~km~s$^{-1}$ (i.e., 2.5 channels) at $\pm 26 \arcsec$ to 
$0.09$ km~s$^{-1}$ (1.3 ch.) at $\pm 52 \arcsec$ in 
C$_3$H$_2$(2$_{12}$-1$_{01}$). In C$^{34}$S(2-1), the velocity shift varies 
from $0.16$ km~s$^{-1}$ (2.6 ch.) at $\pm 26 \arcsec$ to 
$0.14$~km~s$^{-1}$ (2.3 ch.) at $\pm 39 \arcsec$. 
These values translate into a decrease of the C$_3$H$_2$(2$_{12}$-1$_{01}$) 
velocity gradient from 4.8~km~s$^{-1}$~pc$^{-1}$ to 1.3 km~s$^{-1}$~pc$^{-1}$ 
between $26 \arcsec$ and $52 \arcsec$, and to a decrease of the C$^{34}$S(2-1)
velocity gradient from 4.5~km~s$^{-1}$~pc$^{-1}$ to 2.6 km~s$^{-1}$~pc$^{-1}$ 
between $26 \arcsec$ and $39 \arcsec$.  
The decrease of the velocity shift with radius beyond $\pm 26 \arcsec$ 
is {\it in marked contrast with the linear increase expected in the case of 
solid-body rotation}.  We thus conclude that there is strong evidence for differential rotation in the envelope beyond $\sim 3500$~AU.\\
Correcting for inclination and taking all four lines of Fig.~\ref{pvdiag} 
into account, we estimate that the rotational angular velocity decreases by a 
factor of $\sim 5$ from $\Omega = 9 \pm 3$ km~s$^{-1}$~pc$^{-1}$ 
($\sim 3 \times 10^{-13}$ rad~s$^{-1}$) at $r = 2800$~AU ($20 \arcsec$) 
to $\Omega = 1.9 \pm 0.2$~km~s$^{-1}$~pc$^{-1}$ 
($\sim 6 \times 10^{-14}$ rad~s$^{-1}$) at $r = 7000$~AU ($50 \arcsec$).
At a radius $r \sim 11000$~AU, our C$^{18}$O(1-0) and C$^{18}$O(2-1)
observations suggest an even smaller angular velocity, 
$\Omega \simlt 0.5-1$ km~s$^{-1}$~pc$^{-1}$ ($\sim 1.5-3 \times 10^{-14}$ 
rad~s$^{-1}$).
On scales smaller than the beam, the intrinsic rotation velocity pattern 
is uncertain due to insufficient spatial resolution 
(the beam HWHM angular radius is $\sim 13~\arcsec$ at 3 mm, corresponding to
a physical radius of $\sim 1800$ AU).
However, two indirect arguments suggest that the differential rotation
pattern observed here between $\sim 3500$~AU and  $\sim 11000$~AU continues
down to smaller ($\sim 1000-2000$ AU) scales. First, such a differential
rotation pattern, combined with a lower level of molecular 
depletion near envelope center in N$_2$H$^+$ (cf. \S~\ref{obs_width} and 
\S~\ref{molec_depl} below), would explain the higher velocity gradient 
(10~km~s$^{-1}$~pc$^{-1}$) measured in N$_2$H$^+$(101-012) 
over $\pm 20 \arcsec$ compared to the other three lines shown 
in Fig.~\ref{pvdiag}. 
Second, using NH$_3$ interferometric observations sensitive to 
$\sim 5 \arcsec-15 \arcsec $ scales, \citet{Wootten01} have 
recently reported an even larger gradient ($\sim 15$ km~s$^{-1}$~pc$^{-1}$) 
than our present N$_2$H$^+$ value.

%
\subsection{Spectroscopic signature of collapse}
\label{obs_infall}

\begin{figure}
\resizebox{\hsize}{!}{\includegraphics[width=6 cm,angle=270]{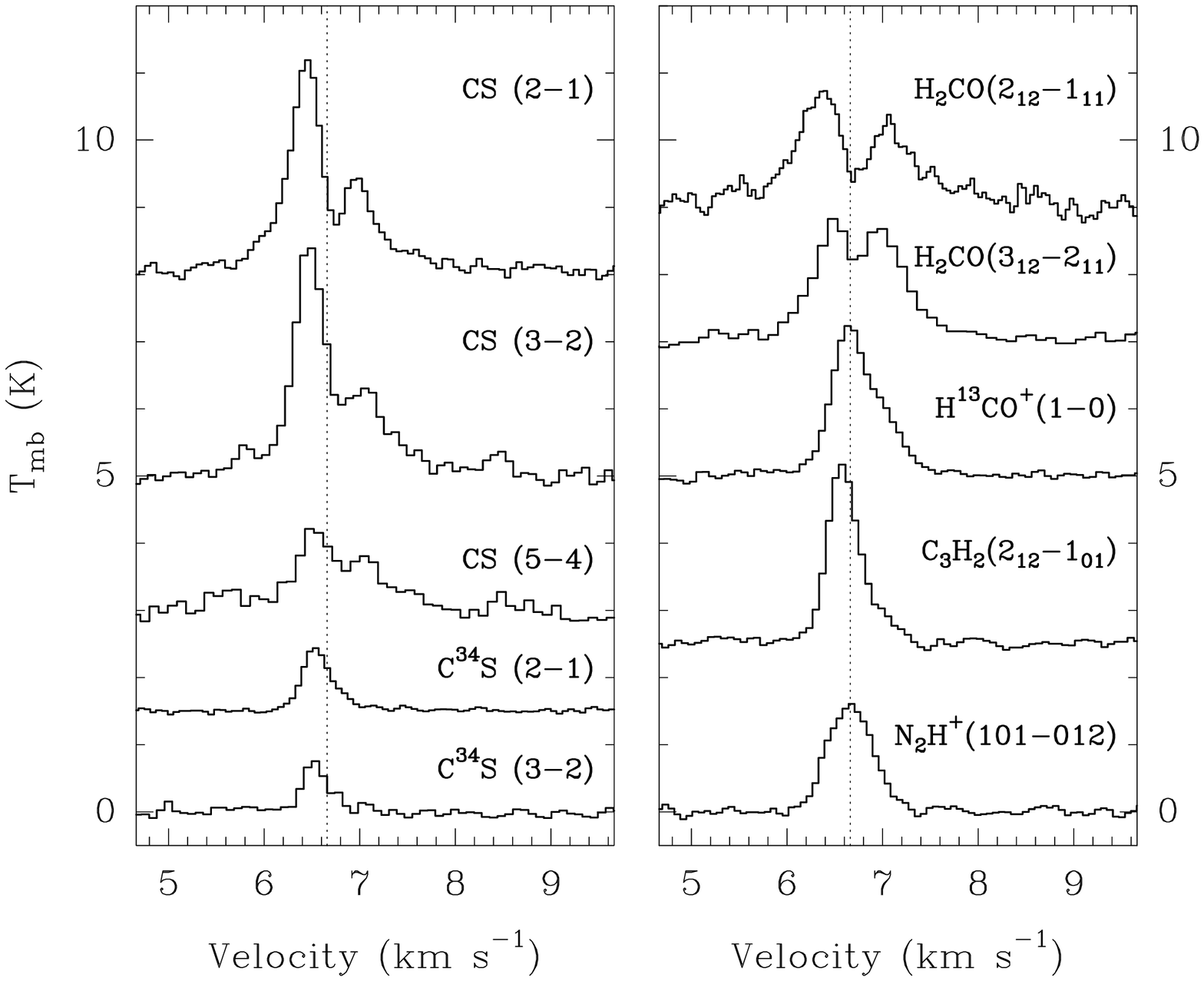}}
\vspace*{-0.5cm}
\caption{Spectral line profiles observed at the central position of 
IRAM~04191 in the optically thick CS(2-1), CS(3-2), CS(5-4), 
H$_2$CO(2$_{12}$-1$_{11}$), H$_2$CO(3$_{12}$-2$_{11}$) lines, and the low 
optical depth C$^{34}$S(2-1), C$^{34}$S(3-2), H$^{13}$CO$^+$(1-0), 
C$_3$H$_2$(2$_{12}$-1$_{01}$), N$_2$H$^+$(101-012) lines. The dotted line 
indicates the source systemic velocity estimated from a 
seven-component hyperfine structure fit to the N$_2$H$^+$(1-0) multiplet.}
\label{spec_center}
\end{figure}

As already pointed out by \citeauthor*{Andre99}, the classical spectroscopic 
signature of infall motions \citep[cf.][]{Evans99,Myers00} is seen toward 
IRAM~04191. Optically thick lines such as CS(2-1), CS(3-2), 
CS(5-4), H$_2$CO(2$_{12}$-1$_{11}$) and H$_2$CO(3$_{12}$-2$_{11}$) are 
double-peaked and skewed to the blue, while low optical depth lines such 
as C$^{34}$S(2-1) and C$^{34}$S(3-2) peak in the dip of the self-absorbed 
lines (see spectra observed at the central position in Fig.~\ref{spec_center}).
Blue-skewed CS(2-1) and CS(3-2) spectra are observed in an extended region, 
up to an angular radius of at least $40 ''$ (5600 AU) from source center  
(see cut taken  
perpendicular to the outflow axis in Fig.~\ref{compare_1d} below).
Such asymmetric line profiles with a blue peak stronger than the red peak 
are expected in a collapsing envelope 
when the line excitation temperature increases toward the center.
There is therefore strong evidence for the presence of extended inward motions
in the IRAM~04191 envelope. 

The blue-to-red peak intensity ratio of the CS lines
is weaker toward the south-east (i.e. the envelope hemisphere red-shifted 
by rotation), while the asymmetry is stronger toward the north-west 
(i.e. the hemisphere blue-shifted by rotation) where the red peak is even 
barely visible. 
This behavior is in qualitative agreement with the expected distortion of the
infall asymmetry due to rotation when the rotation velocity does not dominate
over the infall velocity\footnote{When the rotation 
velocity becomes comparable to the infall velocity, the asymmetry may even be
reversed \citep[e.g.][]{Walker94,Ward01}.} \citep{Zhou95}.

The dips of the optically thick CS(2-1), CS(3-2), and CS(5-4) lines 
have velocities of $6.75 \pm 0.03$ km~s$^{-1}$, $6.81 \pm 0.04$ km~s$^{-1}$, 
and $6.86 \pm 0.05$ km~s$^{-1}$, respectively, and are redshifted relative to 
the source systemic velocity of $6.66 \pm 0.03$~km~s$^{-1}$ (see 
Fig.~\ref{spec_center}). The latter value results from a 7-component Gaussian 
fit to the N$_2$H$^+$(1-0) multiplet, using the hyperfine structure (hfs) 
method of the CLASS reduction software, and assuming the relative frequencies 
and intensities of the 7 hyperfine components determined by \citet{Caselli95},
and the N$_2$H$^+$(101-012) rest frequency of \citet{Lee01} 
(see Table~\ref{tab_freq}). 
The uncertainty on the \citet*{Lee01} frequency is estimated to be  
$\sim 0.02$ km~s$^{-1}$, to which we conservatively add (in quadrature)
an uncertainty of 0.015 km~s$^{-1}$ arising from a maximum
pointing error of $\sim 5\arcsec $, given the velocity gradient discussed 
in \S~\ref{obs_rotation} above. This gives a final uncertainty of 
$\sim 0.03$~km~s$^{-1}$ on the source systemic velocity.
The dips of the self-absorbed CS(2-1), CS(3-2), and CS(5-4) spectra are 
thus redshifted relative to the systemic velocity by $0.09 \pm 0.04$ 
km~s$^{-1}$, $0.15 \pm 0.05$ km~s$^{-1}$, and $0.20 \pm 0.06$ km~s$^{-1}$, 
respectively. These absorption dips are presumably produced 
by the outer layers of the front hemisphere of the envelope.
The fact that they are redshifted provides a second indication that 
inward motions are present in the outer envelope where the opacity of the 
CS lines is of order unity. More quantitatively, the main-beam brightness 
temperature of the dip in the central CS(2-1) spectrum is $\sim 0.7$~K, 
which matches the peak temperature of the spectra taken at $\sim 70 ~\arcsec$ 
from source center. These spectra are still optically thick since we 
measure a CS(2-1) to C$^{34}$S(2-1) integrated intensity ratio of only 
$\sim 5$, i.e., $\sim 4$ times less than the standard CS to C$^{34}$S
isotopic ratio of $\sim 22$ \citep{Wilson94}. 
Assuming a spherically symmetric envelope, we conclude that the 
absorbing shell producing the dip in the central CS(2-1) spectrum has an 
angular radius larger than $\sim 70\arcsec$. 
The observed $\sim 0.1$ km~s$^{-1}$ redshift of the dip is thus suggestive 
of inward motions $\sim 0.1$ km~s$^{-1}$ extending up to a  
radius of at least $10000$ AU.
Radiative transfer simulations confirm this conclusion
(see Sect.~\ref{simul_1d} below).

On the other hand, the CS(5-4) emission is much more concentrated spatially 
than the CS(2-1) emission, and confined to the inner $20 ~\arcsec$ (FWHM)
region (see the non-detection at $10\arcsec$ in Fig.~\ref{compare_1d} below). 
The radius of the 
shell producing the absorption dip in the central CS(5-4) spectrum must 
therefore be smaller than $10 ~\arcsec$. 
The observed $\sim 0.2$~km~s$^{-1}$ redshift of the CS(5-4) dip 
then suggests faster inward motions in the central ($r < 1400$ AU) region.

%
\subsection{Linewidths: Evidence for turbulent infall ?}
\label{obs_width}

The C$^{34}$S(2-1) and C$^{34}$S(3-2) spectra displayed 
in Fig.~\ref{spec_center} 
are slightly asymmetric and skewed to the blue, which suggests  
they are marginally optically thick and showing some infall asymmetry 
\citep[see, e.g., Fig.~1 of][]{Myers95}. We measure (FWHM) linewidths of $0.38 
\pm 0.01$ km~s$^{-1}$ for C$^{34}$S(2-1) and $0.32 \pm 0.03$ km~s$^{-1}$ for 
C$^{34}$S(3-2), which are 3.8 and 3.2 times larger than the 
thermal broadening of C$^{34}$S at a kinetic temperature 
T$_\mathrm{K} = 10$~K, respectively 
(see \S~\ref{mass_tk} for constraints on T$_\mathrm{K}$). Radiative transfer 
simulations (cf. Sect.~\ref{simul_1d}) indicate opacities of $\sim 1.5$ and 
$\sim 1$ for C$^{34}$S(2-1) and C$^{34}$S(3-2), respectively. 
Therefore, line saturation effects cannot 
broaden the C$^{34}$S(2-1) and C$^{34}$S(3-2) spectra by more than  
$\sim 30 \%$ and $\sim 20 \%$, respectively, and the linewidths are primarily
nonthermal. Motions such as infall, 
rotation, outflow, or ``turbulence'' along the line of sight are  
required to explain such nonthermal linewidths. The nonthermal motions do not
dominate over thermal motions, however, since the C$^{34}$S(2-1) 
and C$^{34}$S(3-2) linewidths represent only $\sim 80 \%$ of 
the thermal velocity dispersion for a mean particle of molecular weight
$\mu =2.33$. The N$_2$H$^+$(101-012) line is slightly broader, with a 
FWHM $\sim 0.55$ km~s$^{-1}$, i.e., 1.2 times broader than the (mean particle)
thermal velocity dispersion and $\sim 1.5$ times larger than the C$^{34}$S 
linewidths. The hyperfine structure fit to the N$_2$H$^+$ multiplet 
(see \S~\ref{obs_infall}) yields an optical depth of $\sim 0.85$ for the 
isolated N$_2$H$^+$(101-012) component, suggesting negligible ($\simlt 15\% $)
optical depth broadening. The level of optical depth broadening 
should thus be more pronounced in C$^{34}$S and cannot explain the difference 
in linewidth between N$_2$H$^+$ and C$^{34}$S.

We propose that this difference in linewidth between N$_2$H$^+$ and C$^{34}$S 
results from a combination of higher infall/rotation velocities and 
lower N$_2$H$^+$ depletion toward the center. \citet{Bergin97} have shown that
sulphur-bearing molecules such as CS are strongly depleted when the density 
increases, whereas N$_2$H$^+$ 
remains in the gas phase, at least up to densities n$_{\mbox{\tiny H}_2} 
\simlt 10^6$ cm$^{-3}$. Indeed, we measure a decrease of 
the C$^{34}$S(2-1)/N$_2$H$^+$(101-012) integrated intensity ratio 
by a factor of 2 from $\sim 5000$ AU to $\sim 2000$ AU. 
As both lines have nearly the same critical density and are 
approximately optically thin\footnote{Based on the opacities derived above,
optical depth effects cannot account for more than a factor of 1.3 
decrease in the C$^{34}$S(2-1)/N$_2$H$^+$(101-012) ratio.},
the decrease of the integrated intensity ratio may be interpreted as a 
decrease of the C$^{34}$S/N$_2$H$^+$ abundance ratio toward the center.

Finally, we note that the N$_2$H$^+$(101-012) linewidth peaks at the 
central position (as shown by a linewidth-position plot along the 
direction perpendicular to the ouflow axis).
N$_2$H$^+$ may thus be more sensitive 
to higher velocity material produced by, e.g., infall, rotation, 
or outflow near the central protostellar object. 
As N$_2$H$^+$ is generally underabundant in molecular outflows 
\citep[e.g.][]{Bachiller97}, the central broadening 
of N$_2$H$^+$(1-0) is most likely due to infall and/or rotation motions.

%
\subsection{The CS line wing emission}
\label{obs_wing}

\begin{figure}
\resizebox{\hsize}{!}{\includegraphics[width=6. cm,angle=270]{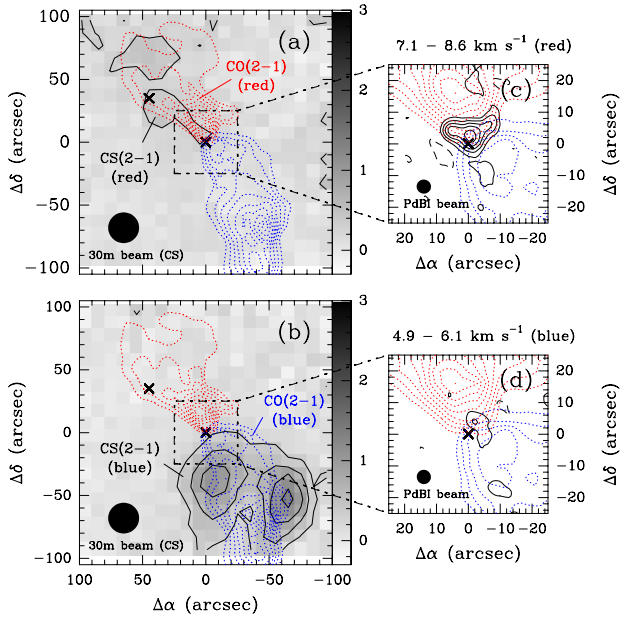}}
\caption{CS(2-1) integrated intensity maps taken with the 30m telescope (left)
and, on a smaller scale, with the PdBI interferometer (right). 
(a) Single-dish map integrated in the 7.1 to 8.6 km~s$^{-1}$ velocity range 
(red), overlaid on the CO(2-1) outflow contours from \citeauthor*{Andre99}. 
(b) Same as (a) for the 4.6 to 6.1 km~s$^{-1}$ velocity range (blue). 
The CS wing emission ({\it solid contours and gray scale}) appears 
to be correlated with the CO outflow emission ({\it dashed contours}). 
The two crosses mark the central positions of IRAM~04191 [at (0,0)] 
and IRAS~04191. The CS(2-1) contour step is 0.3 K~km~s$^{-1}$; the gray scale
(shown on the right of each map) ranges from $-0.3$ to 3~K~km~s$^{-1}$. 
(c) Cleaned interferometer map, integrated over the 7.1 to 8.6 km~s$^{-1}$ 
velocity range (red), overlaid on the CO(2-1) contours of the outflow. (d) 
Same as (c) for the 4.9 to 6.1 km~s$^{-1}$ velocity range (blue). The CS(2-1) 
contour step is 0.25 Jy~beam$^{-1}$~km~s$^{-1}$. The clean beam is shown on 
the bottom left. Note how the CS(2-1) red wing emission detected on a 
$\sim 5\arcsec$ scale coincides with the base of the red lobe of the CO 
outflow, while the CS(2-1) blue wing emission is almost entirely resolved out 
by the interferometer.}
\label{wing_map}
\end{figure}

The morphology of 
the single-dish CS(2-1) maps integrated over the 4.6--6.1~km~s$^{-1}$ (blue) 
and 7.1--8.6 km~s$^{-1}$ (red) velocity ranges strongly suggests  
that the CS line wing emission is dominated by material associated with 
the outflow (see Fig.~\ref{wing_map}a \& Fig.~\ref{wing_map}b).  
Some CS(2-1) emission is detected at the edges of both the red and the blue
lobe of the CO outflow: the CS(2-1) red wing is 
relatively weak and concentrated at the south-eastern edge and the tip of the 
red CO lobe (Fig.~\ref{wing_map}a), while the CS(2-1) blue wing 
is much stronger and distributed in two spots on either side of the blue CO 
lobe (Fig.~\ref{wing_map}b). 
This ``high-velocity'' CS(2-1) emission is likely to arise from dense, shocked 
material entrained by the outflow. 

The redshifted CS(2-1) emission detected on smaller scales by the PdBI 
interferometer in the 7--8 km~s$^{-1}$ velocity range also appears to be
associated with the outflow. This emission arises from the base of the 
red CO lobe and its shape closely follows the edges of the outflow lobe 
(see Fig.~\ref{wing_map}c). No emission was detected by PdBI 
in the blue wing range (see Fig.~\ref{wing_map}d) or at the source systemic 
velocity. The blueshifted CS(2-1) emission seen in the single-dish map
(Fig.~\ref{wing_map}b) thus appears to be more extended than the redshifted  
emission and is likely resolved out by the interferometer.  
Such a difference in spatial extent between blueshifted and redshifted 
emission, opposite to what infall motions would produce, is expected if the 
CS line wing emission arises from outflowing material. 
Likewise, the emission detected by the 30m telescope near the systemic 
velocity, i.e., close to the dip of the CS(2-1) line 
(Fig.~\ref{spec_center}), arises from extended foreground material on 
scales $\sim 70\arcsec $ (see \S~\ref{obs_infall}) and is also resolved out 
by PdBI.

\subsection{Low degree of ionization}
\label{obs_ionization}

Following, e.g., \citet{Caselli98}, we can estimate the degree of 
ionization in the IRAM~04191 envelope from the observed value of the 
abundance ratio R$_\mathrm{D} = [\mathrm{DCO}^+]/[\mathrm{HCO}^+]$.  
We measure an integrated intensity ratio
$I_{\mathrm{DCO}^+}/I_{\mathrm{H}^{13}\mathrm{CO}^+} = 3.3 \pm 0.6$ at  
the central position. Assuming optically thin H$^{13}$CO$^+$(3-2) and 
DCO$^+$(3-2) emission, this implies  
$[\mathrm{DCO}^+]/[$H$^{13}$CO$^+$$] = 3.3 \pm 0.6$. Adopting 
a $[^{12}\mathrm{C}]/[^{13}\mathrm{C}]$ abundance ratio of 77 in the local ISM 
\citep{Wilson94}, we thus derive R$_\mathrm{D} = 0.04 \pm 0.01$, which is 
identical to the ratio measured by \citet{Caselli01b} toward the 
central position of L1544.  
As the same density ($\sim 10^6$ cm$^{-3}$) is probed by the observations in 
both cases, we conclude that the ionization degree $x_i \sim 2 \times 
10^{-9}$ derived in L1544 by \citet{Caselli01b} based on their chemical models 
should be representative of the ionization degree in the IRAM~04191 envelope 
at this density (assuming similar depletion factors for CO, which seems 
likely -- see \S~\ref{molec_depl} below). 

\section{Radiative transfer modeling: 1D spherical simulations}
\label{simul_1d}

Here, we use the radiative transfer code MAPYSO \citep{Blinder97} to model 
the observed line spectra and set quantitative constraints on the 
kinematics of the IRAM 04191 envelope (see Appendix for details about 
the code). 
For simplicity, we use a piecewise powerlaw description for the spatial
variations of the kinetic temperature, density, molecular abundance, 
infall velocity, and rotational velocity (in Sect.~\ref{simul_2d} below) 
across the envelope (see Fig.~\ref{model_1d}).

%
\subsection{Model inputs: mass distribution and kinetic temperature profile}
\label{mass_tk}

\begin{figure}
\resizebox{\hsize}{!}{\includegraphics[angle=270]{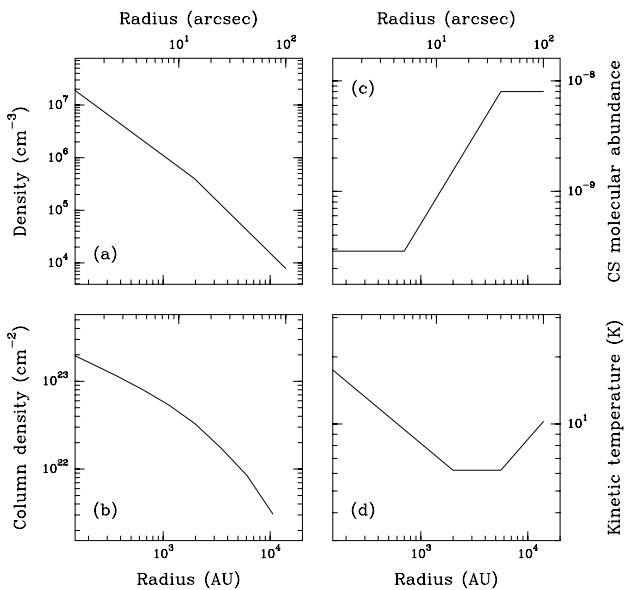}}
\caption{Input physical properties of the 1D spherical envelope model 
used to fit the observed CS and C$^{34}$S spectra 
(see Fig.~\ref{compare_1d}): Plots of density (a), column density (b), 
CS/H$_2$ relative abundance (c), and kinetic temperature (d) as a function of 
distance from envelope center.}
\label{model_1d}
\end{figure}

We use the envelope mass distribution derived by \citet{Motte01} 
(hereafter \citeauthor*{Motte01}) from 1.3~mm dust continuum observations 
with the 30m telescope. \citeauthor*{Motte01} estimate that the envelope 
mass contained within a radius of 
4200~AU is $M_\mathrm{env}(r < 4200~ \mathrm{AU}) = 0.45$ M$_\odot$ with an 
uncertainty of a factor of 2 on either side, assuming a mean dust temperature 
$T_\mathrm{dust} = 12.5 \pm 2.5$ K obtained by \citeauthor*{Andre99} from a 
graybody fit to the $\lambda > 90$ $\mu$m portion of the SED (see Fig.~3 of
 \citeauthor*{Andre99}). In addition, the radial structure analysis of 
\citeauthor*{Motte01} 
indicates an average radial intensity profile $I \propto \theta^{-m}$ with 
$m = 0.6 \pm 0.1$ in the range of angular radii $\theta \sim 11''-100''$ 
(i.e., $r \sim $~1500-14000 AU). Assuming a dust temperature profile 
$T_\mathrm{dust} \propto r^{-q}$ with $q = -0.2 \pm 0.2$, 
\citeauthor*{Motte01} obtain a density profile $\rho \propto r^{-p}$ with
$p = m+1-q = 1.8 \pm 0.3$. Given the low bolometric luminosity of IRAM~04191, 
the dust temperature
is indeed expected to rise outward ($q < 0$) due to external heating by  
the interstellar radiation field \citep[e.g.][]{Masunaga00,Evans01,Zucconi01}. 
The regime of central heating by the accreting protostar ($q \sim 0.4$)
is likely confined to the inner $\sim 2000$~AU radius region  (cf. 
\citeauthor*{Motte01}). Since a good thermal coupling between gas and dust 
grains is expected for densities $n_{\mbox{\tiny H$_2$}} > 10^5$ cm$^{-3}$ 
\citep[e.g.][]{Ceccarelli96,Doty97}, the gas kinetic temperature 
profile should track the dust temperature profile, at least up to a radius of 
$\sim 4000$ AU.

We also have some constraints on the gas kinetic temperature from 
our C$^{18}$O, N$_2$H$^+$, and CS observations. We measure a C$^{18}$O(2-1) to
C$^{18}$O(1-0) integrated intensity ratio 
$I_{\mbox {\tiny C$^{18}$O(2-1)}}/I_{\mbox {\tiny C$^{18}$O(1-0)}} = 1.5 \pm 
0.3$ in the range of angular radii 
$0''-60''$, using main beam temperatures and after degrading the resolution 
of the $J= 2-1$ data to that of the $J= 1-0$ data.
As the critical densities 
of C$^{18}$O(1-0) and C$^{18}$O(2-1) are $3 \times 10^3$~cm$^{-3}$ and 
$2 \times 10^4$~cm$^{-3}$ at 10~K, respectively, both lines should be 
thermalized in most of the envelope. Assuming local thermodynamic equilibrium
(LTE) and optically thin emission, we derive an 
excitation temperature $T_{\mbox{\tiny ex}}(\mbox{C$^{18}$O}) = 10 \pm 2$ K, 
which should be a good estimate of the gas kinetic temperature in 
the low-density outer ($r \sim 6000$~AU) part of the envelope probed 
by C$^{18}$O. 
Likewise, the excitation temperature of the dense-gas tracer
N$_2$H$^+$(101-012) may be estimated from the relative 
intensities of the seven components of the N$_2$H$^+$(1-0) multiplet, assuming
the same excitation temperature for all components 
(cf. \S~\ref{obs_infall}).
This method
yields $T_{\mbox{\tiny ex}}(\mbox{N$_2$H$^+$}) = 5.5 \pm 0.5$ K. 
With a critical density $\sim 2 \times 10^5$ cm$^{-3}$, the 
N$_2$H$^+$(1-0) multiplet is probably thermalized only in the inner 
($r < 2000$~AU) envelope. We thus obtain a lower limit of $\sim 6$~K 
for the gas kinetic temperature in the dense 
($n_{\mbox{\tiny H$_2$}} > 10^5$ cm$^{-3}$), inner part of the envelope. 
Finally, the weak intensities of the optically thick CS(2-1) and CS(3-2) 
lines require a low gas kinetic temperature $\sim$~6-7 K in the range 
of radii $2000 - 6000$ AU. Given the 
density profile shown in Fig.~\ref{model_1d}a, a uniform gas temperature of 
10~K would produce CS(2-1) and CS(3-2) spectra with main beam temperatures 
about 1-2 K stronger than the observed temperatures. 
Beyond $r \sim 6000$~AU, the gas temperature is likely to increase 
to the typical $\sim 10$~K temperature of the Taurus cloud 
\citep[see, e.g.,][]{Benson89}.
However, the gas temperature profile in the outer parts of the envelope 
has little influence on the CS and C$^{34}$S spectra since the observed
lines are far from LTE there.

In summary, the gas kinetic temperature profile is likely to present a minimum
of $\sim 6$ K at $\sim 10\arcsec-20\arcsec$ (i.e., $r \sim $~1400-2800 AU) and 
to reach a value of $\sim 10$ K in the outer parts of the envelope.

%
\subsection{Molecular depletion}
\label{molec_depl}

Assuming a standard isotopic ratio 
$\chi_{\mbox{\tiny CS}}/\chi_{\mbox{\tiny C$^{34}$S}} = 22$, 
a $[\mathrm{CS}]/[\mathrm{H}_2]$ abundance ratio of $8 \times 10^{-9}$ is
required to match the C$^{34}$S(2-1) integrated intensity at an angular radius 
$40~\arcsec-50~\arcsec$. 
But a uniform abundance with such a value produces too strong C$^{34}$S(2-1),  
C$^{34}$S(3-2), and CS(5-4) spectra toward the center. A good fit to the 
C$^{34}$S(2-1) integrated line intensities is obtained by assuming that 
the relative CS abundance drops 
by a factor of $\sim 20$ toward the center (see also \S~\ref{obs_width}).
Such a depletion factor for CS is comparable to those observed in starless 
cores such as L1544 \citep[e.g.][]{Tafalla02}.
%

\subsection{Two regimes of infall}
\label{analyze_1d}

\begin{figure*} [!ht]
\centering
\includegraphics[width=15cm,angle=0]{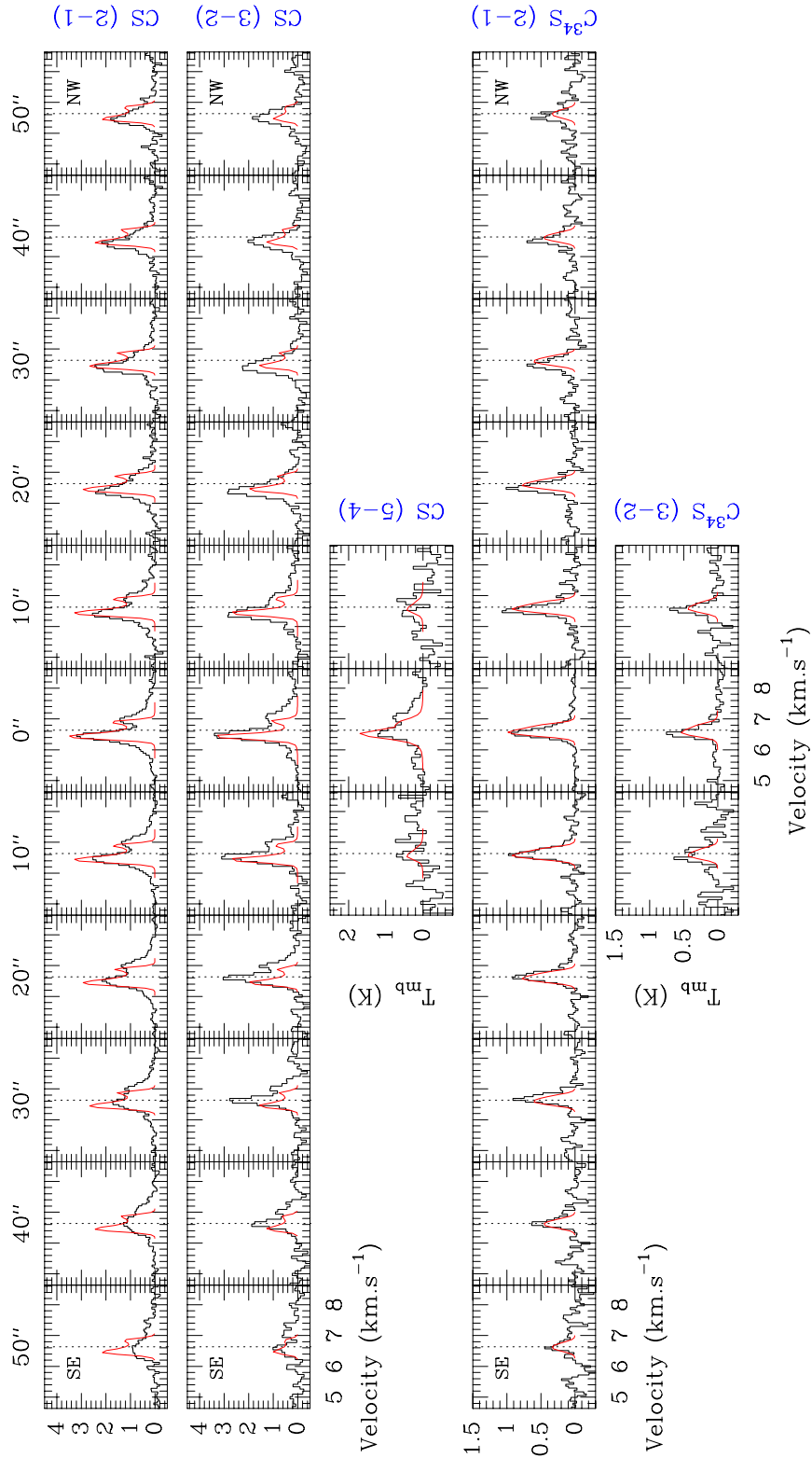}
 \caption{CS(2-1), CS(3-2), CS(5-4), C$^{34}$S(2-1), and C$^{34}$S(3-2) 
spectra (in units of main beam temperature) observed with the 30m telescope
along the direction perpendicular to the outflow axis (histograms). 
The dotted line indicates our best-fit estimate (6.63 km~s$^{-1}$) of the 
envelope systemic velocity based on our CS/C$^{34}$S modeling.
Synthetic spectra corresponding to the `best-fit' 1D spherical collapse model 
described in \S~\ref{analyze_1d} (cf. Fig.~\ref{model_1d} and
Fig.~\ref{model_vel}a,b for model parameters) are superposed.}
 \label{compare_1d}
\end{figure*}

\begin{figure*} [!ht]
\centering
\resizebox{\hsize}{!}{\includegraphics[width=9cm,angle=270]{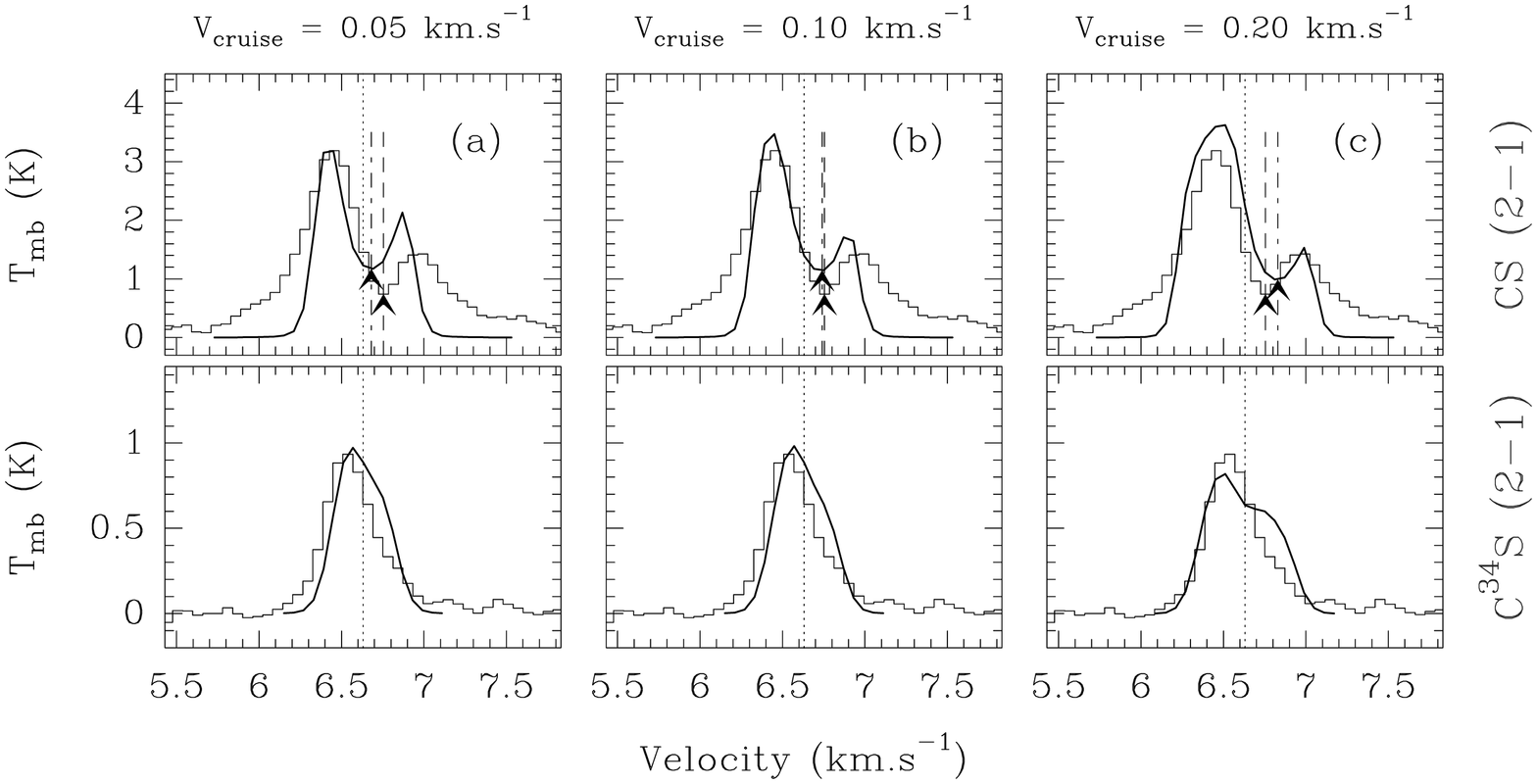}}
 \vspace*{-0.6cm}
 \caption{Influence of the outer infall velocity field on the CS(2-1) 
({\it upper row}) and C$^{34}$S(2-1) ({\it lower row}) line profiles. 
Synthetic spectra (solid curves) 
corresponding to a range of model infall velocities are superposed 
on the spectra observed at the central position of IRAM 04191 (histograms). 
All models have $v_{inf} = v_0 = 0.2$ km~s$^{-1}$ at $r_0 = 750$~AU, 
$v_{inf} \propto r^{-0.5}$ for $r < r_{break}$, and 
$v_{inf} = v_{cruise}$ for $r > r_{break}$. 
The models differ by the value of $r_{break}$,  
or equivalently by the value of the infall velocity at large radii, 
$v_{cruise} \equiv v_0 \times \left(\frac{r_{break}}{r_0}\right)^{-0.5} $:
$r_{break} = 12000$ AU (i.e., $v_{cruise} = 0.05$ km~s$^{-1}$) in (a), 
$r_{break} = 3000$ AU (i.e., $v_{cruise} = 0.1$ km~s$^{-1}$) in (b) 
($\equiv $ preferred model shown in Fig.~\ref{compare_1d}), and 
$r_{break} = 750$ AU (i.e., $v_{cruise} = 0.2$ km~s$^{-1}$) in (c). 
The dotted vertical line marks the envelope systemic velocity 
as in Fig.~\ref{compare_1d}, while  
the dashed and dash-dotted lines mark the velocities of the dips in the
observed and synthetic CS(2-1) spectra, respectively. 
Note that the outer infall velocity $v_{cruise}$ is too small in model (a)
and too large in model (c) to match the observed redshift of the CS(2-1) dip
and the width of the C$^{34}$S(2-1) line.}
 \label{zoom_vinfext}
\end{figure*}

In Fig.~\ref{compare_1d}, we present a series of synthetic spectra emitted 
by a spherically symmetric model envelope with the input structure 
shown in Fig.~\ref{model_1d} and our best estimate of the infall velocity 
field (shown by the solid line in Fig.~\ref{model_vel}a below).
The model spectra are overlaid on the 
multitransition CS and C$^{34}$S spectra observed along the direction 
perpendicular to the outflow axis, so as to minimize the effects of the 
outflow. The blue infall asymmetry of the model optically thick lines and 
the position of the CS(2-1) dip match the observations well. 
The widths of the optically thin lines are also well reproduced. 
The main shortcoming of the model is that it does not reproduce
the fairly strong emission present in the wings of the observed CS spectra.
As discussed in  \S~\ref{obs_wing}, we attribute these wings 
to the fraction of envelope material entrained by the outflow. 

\begin{figure*}[!ht]
\centering
\resizebox{\hsize}{!}{\includegraphics[width=9cm,angle=270]{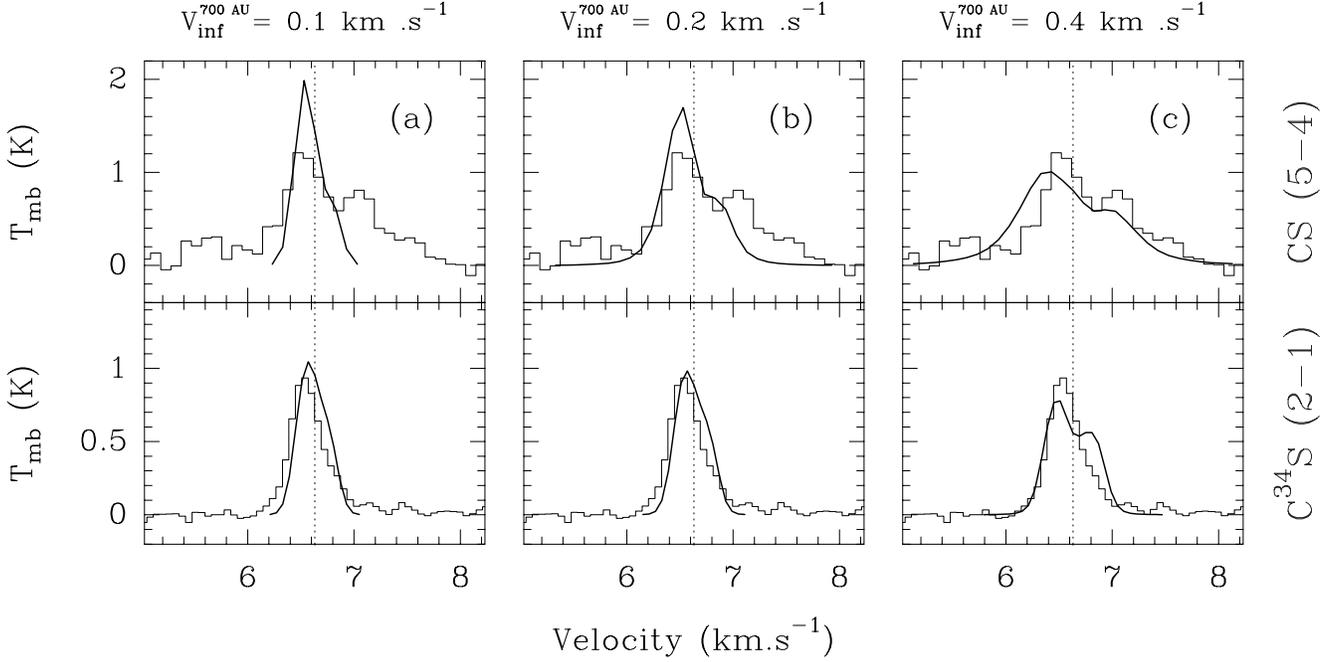}}
 \vspace*{-0.6cm}
 \caption{Influence of the inner infall velocity field on the CS(5-4) 
({\it upper row}) and C$^{34}$S(2-1) ({\it lower row}) line profiles. 
Synthetic spectra (solid curves) 
corresponding to a range of model infall velocities are superposed 
on the spectra observed at the central position of IRAM 04191 (histograms). 
All models have $v_{inf} \propto r^{-0.5}$ for $r < r_{break}$ 
and $v_{inf} = 0.1$ km~s$^{-1}$ for $r > r_{break}$. 
The models differ by the value of $r_{break}$,  
and therefore by the value of the infall velocity at the 700~AU radius of the
CS(5-4) beam:
$r_{break} = 0$ AU (i.e., $v_\mathrm{700 AU} = 0.1$ km~s$^{-1}$) in (a), 
$r_{break} = 3000$ AU (i.e., $v_\mathrm{700 AU} = 0.2$ km~s$^{-1}$) in (b) 
($\equiv $ preferred model shown in Fig.~\ref{compare_1d}), and 
$r_{break} = 13900$ AU (i.e., $v_\mathrm{700 AU} = 0.4$ km~s$^{-1}$) in (c). 
The dotted vertical line marks the envelope systemic velocity as in 
Fig.~\ref{compare_1d}.
Note that the CS(5-4) linewidth is too small in model (a) and the 
C$^{34}$S(2-1) linewidth is too large in model (c), 
suggesting $0.1 < v_{inf} < 0.4$~km~s$^{-1}$ at $r \sim 700$ AU.}
 \label{zoom_vinfint}
\end{figure*}

\begin{figure*}[!ht]
\centering
\resizebox{\hsize}{!}{\includegraphics[width=9cm,angle=270]{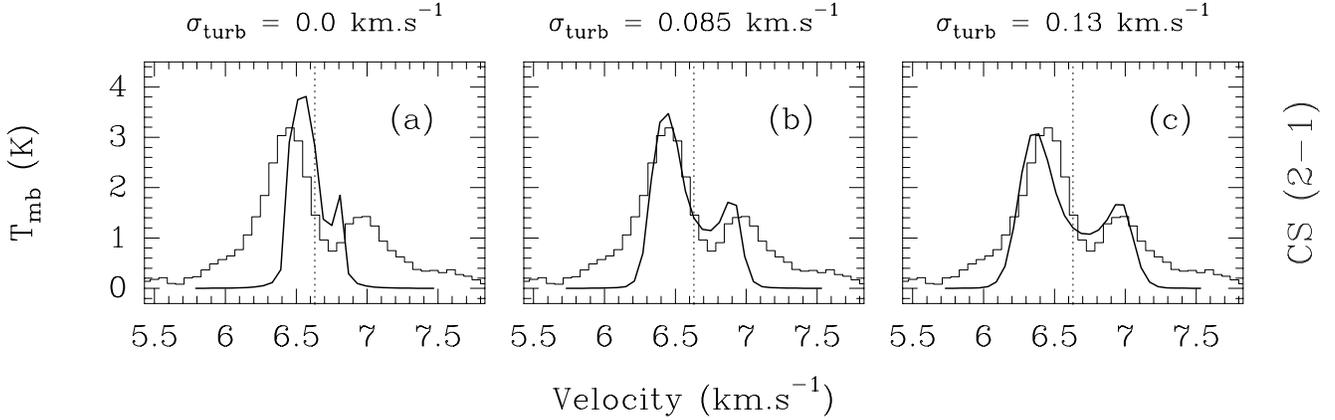}}
 \vspace*{-0.6cm}
 \caption{Influence of the turbulent velocity field on the CS(2-1) 
spectra. The line profiles observed at the central 
position of IRAM 04191 (histograms) are compared to 
synthetic spectra (solid lines) corresponding to models with a 
uniform turbulent velocity for various values 
of the velocity dispersion $\sigma_{turb}$: (a) 
$\sigma_{turb} = 0$ km~s$^{-1}$, (b) $\sigma_{turb} = 0.085$ km~s$^{-1}$  
($\equiv $ reference model shown in 
Fig.~\ref{compare_1d}), (c) $\sigma_{turb} = 0.13$ km~s$^{-1}$. 
The dotted vertical line marks the envelope systemic velocity 
as in Fig.~\ref{compare_1d}.
Note that the dip of the model CS(2-1) spectrum is too narrow in (a) and 
too broad in (c).}
 \label{zoom_vturb}
\end{figure*}

\begin{figure*}[!ht]
\resizebox{\hsize}{!}{\includegraphics[angle=270]{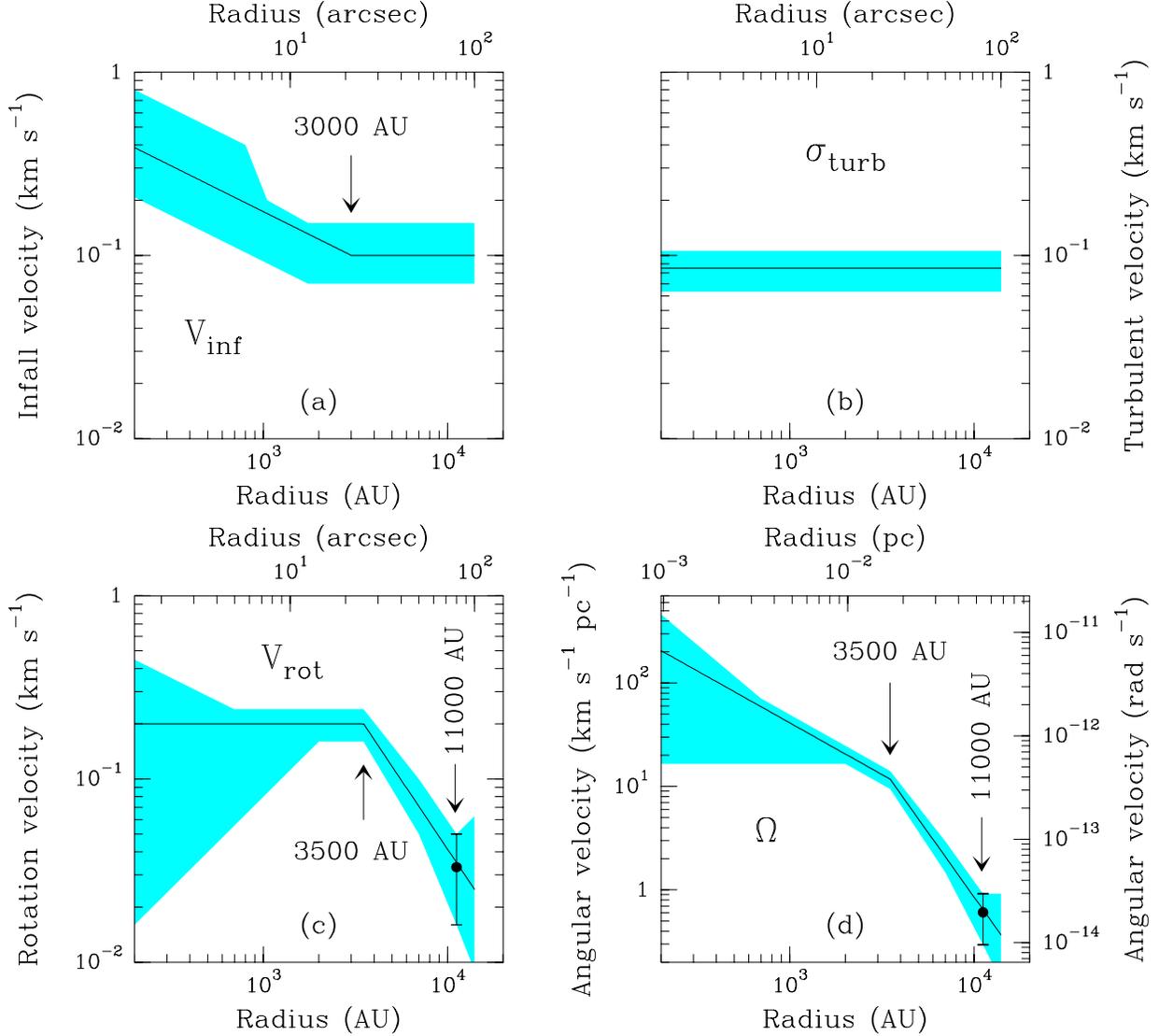}}
\caption{Infall (a), turbulence (b), and rotation (c) velocity fields 
inferred in the IRAM~04191 envelope based on our 1D (\S~\ref{simul_1d})
and 2D (\S~\ref{simul_2d}) radiative transfer modeling. 
The shaded areas show the estimated domains where the 
models match the CS and C$^{34}$S observations reasonably well.
In (a) and (b), the solid lines show the infall velocity and the 
turbulent velocity dispersion in both the 1D and 2D models
(cf. Fig.~\ref{compare_1d} and Fig.~\ref{compare_2d}, respectively)
as a function of radius from envelope center.
In (c), the solid line represents the profile of the 
azimuthal rotation velocity in the 2D envelope model 
(cf. Fig.~\ref{compare_2d}) as a function of  
distance from the outflow/rotation axis. The point with error bar at 11000~AU
corresponds to the velocity gradient observed in C$^{18}$O (cf. 
\S~\ref{obs_rotation}). 
Panel (d) shows the corresponding angular velocity profile.}
\label{model_vel}
\end{figure*}

In order to determine the range of input model parameters that yield 
reasonably good fits to the observed CS and C$^{34}$S spectra, we have 
performed a comprehensive exploration of the parameter space, as illustrated
in Figs.~\ref{zoom_vinfext}, \ref{zoom_vinfint} and \ref{zoom_vturb}.
Strong constraints on the infall velocity arise from the small 
optical depth lines, i.e., C$^{34}$S(2-1) and C$^{34}$S(3-2). The  
widths of these lines set firm upper limits to the absolute value  
of the infall velocity on the size scale of the beam.
We obtain $v_\mathrm{inf} \leq 0.15$ km~s$^{-1}$ at $\sim 1750$ AU and 
$v_\mathrm{inf} \leq 0.2$ km~s$^{-1}$ at $\sim 1150$ AU from C$^{34}$S(2-1) 
and C$^{34}$S(3-2), respectively. 
On the other hand, the amplitude of the blue asymmetry seen in the 
self-absorbed CS(2-1) and CS(3-2) lines, as well as the redshifted position of
the corresponding absorption dips (see \S~\ref{obs_infall}),  
both require a relatively flat, extended infall velocity field 
with $v_\mathrm{inf} \sim 0.10 \pm 0.05$ km~s$^{-1}$ up to 
$r \sim 10000-12000$ AU. The latter value approximately corresponds to 
the radius where the bulk of the absorption occurs in CS(2-1) and CS(3-2)
(see \S~\ref{obs_infall}). 
These constraints are illustrated in 
Fig.~\ref{zoom_vinfext} which shows the effect of varying the infall velocity
field on the central CS(2-1) and C$^{34}$S(2-1) spectra. Three models are
compared: the preferred model displayed in Fig.~\ref{compare_1d} 
is shown in the central panel (Fig.~\ref{zoom_vinfext}b), 
while models with lower and higher infall velocities in the outer part of
the envelope are shown in the left (a) and right (c) panels, respectively.
It can be seen that the position of the CS(2-1) absorption dip
is not redshifted enough in model (a) and too redshifted in model (c) to 
match the observations.
Furthermore, the CS(2-1) blue-to-red asymmetry is too weak 
in model (a) and the C$^{34}$S(2-1) line becomes too broad in model (c) 
compared to the observations.
Only model (b) approximately reproduces the observed position of the CS(2-1) 
dip and the width of the C$^{34}$S(2-1) line.

In the context of a pure infall model, larger velocities 
($v_\mathrm{inf} \sim 0.2-0.4 $ km~s$^{-1}$)
in the inner ($r \sim 700$ AU) part of the envelope 
are suggested by the broad linewidth of the central CS(5-4) spectrum.
This is shown in Fig.~\ref{zoom_vinfint} which compares three models
differing in the magnitude of their infall velocity at 700~AU.
A similar trend is indicated by the broadening of the N$_2$H$^+$(101-012) 
line toward the center (cf. \S~\ref{obs_width}).
However, the central CS(5-4) spectrum 
(which does show infall asymmetry -- cf. Fig.~\ref{spec_center}) may be partly
contaminated by small-scale structure in the outflow.
In particular, the redshifted portion of the CS(5-4) spectrum, 
not reproduced by the model of Fig.~\ref{compare_1d}, 
may be related to the redshifted CS(2-1) emission detected by 
the interferometer on small scales (see \S~\ref{obs_wing}). 
Such a switch between infall-dominated CS emission on large scales and 
outflow-dominated CS emission on small scales is also observed in the Class~0 
object B335 \citep[cf.][]{Wilner00}. 

In addition to infall, a ``turbulent'' velocity field is needed to match the 
width of the dip, and consequently the velocity difference between the blue 
and the red peak, in the optically thick CS(2-1) and CS(3-2) spectra. 
Assuming a uniform turbulent velocity dispersion 
for simplicity, a good compromise between the upper limit set by the 
linewidth of the 
optically thin C$^{34}$S(2-1) and C$^{34}$S(3-2) spectra and the lower limit 
set by the width of the CS(2-1) and CS(3-2) dips is obtained for 
$\sigma_\mathrm{turb} = 0.085 \pm 0.02 $ km~s$^{-1}$ 
(cf. Fig.~\ref{zoom_vturb}). This is equivalent to  
$\Delta v_\mathrm{turb}^\mathrm{FWHM} = \sigma_\mathrm{turb} \times 
\sqrt{8\,\rm{ln}2} = 0.20 \pm 0.05 $ km~s$^{-1}$ and corresponds to  
only half the thermal broadening of the mean molecular particle at 10~K, 
showing that the IRAM~04191 envelope is ``thermally-dominated'' (see also 
\S~\ref{obs_width}) as are Taurus dense cores in general 
\citep[e.g.][]{Myers99}.

The main conclusions of our 1D exploration of the parameter space are 
summarized in Fig.~\ref{model_vel}a and Fig.~\ref{model_vel}b, 
where the shaded areas represent the ranges of infall velocities (a) and 
turbulent velocity dispersion (b) for which acceptable fits are found.
Two infall regimes seem to stand out in Fig.~\ref{model_vel}a: 
the infall velocity is relatively large ($v_\mathrm{inf} \simgt
0.2$~km~s$^{-1}$, supersonic) and increases toward the center 
for $r \simlt 2000-3000$ AU, while it is smaller and roughly 
uniform at $v_\mathrm{inf} \sim 0.10 \pm 0.05$ km~s$^{-1}$
between $\sim 2000-3000$ AU and $\sim 10000-12000$ AU. 
Given the density profile of Fig.~\ref{model_1d}a, such an infall velocity 
field implies a mass infall rate of 
$\dot{M}_\mathrm{inf} \sim 3 \times 10^{-6}$~M$_\odot$~yr$^{-1}$ at 
$r = 1750$ AU. (The density and velocity profiles shown in 
Fig.~\ref{model_1d}a and Fig.~\ref{model_vel}a are such that
$\dot{M}_\mathrm{inf} $ is roughly independent of radius.)
Inside the $r \sim 11000 $~AU region (where non-zero inward motions are 
inferred), the fraction of envelope mass with supersonic 
($\simgt 0.16-0.2$~km~s$^{-1}$) infall velocities is estimated to be only 
$\sim 1-10\%$, depending on the exact value of the sound speed and exact form
of the infall profile (see Fig.~\ref{model_vel}a).

\section{Radiative transfer modeling: Simulations with infall and 
rotation}
\label{simul_2d}

\begin{figure} [!ht]
 \resizebox{\hsize}{!}{\includegraphics[angle=270]{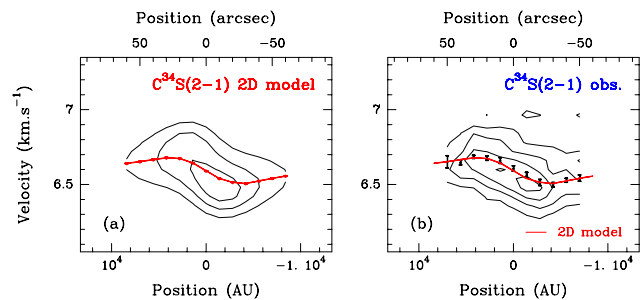}}
 \caption{(a) C$^{34}$S(2-1) position-velocity diagram of the `best-fit'
2D collapse model including differential rotation, taken along the 
direction perpendicular to the rotation axis. 
(b) Observed C$^{34}$S(2-1) position-velocity diagram along the 
direction perpendicular to the outflow axis.
In both diagrams, the dots 
represent the peak velocities 
derived from Gaussian fits at angular radii larger than $20 ~\arcsec$ and from 
centroid (first moment) estimates for radii smaller than $20 ~\arcsec$; 
the curve shows the variation of the peak velocity (estimated in the same 
way) in the model.
}
 \label{pvdiagmod}
\end{figure}

%
\subsection{Quasi 2D simulations}
\label{mapyso_2d}

To account for the effects of rotation in the envelope 
(see \S~\ref{obs_rotation} and Fig.~\ref{pvdiag}), we have performed 
``quasi''-2D simulations with the following approximation. The non-LTE 
level populations are still calculated with a 1D Monte Carlo method (see 
Appendix) assuming a spherical envelope with the same characteristics 
as the model described in \S~\ref{analyze_1d}. We then add a 
cylindrical rotation velocity field to the 1D model of 
\S~\ref{analyze_1d} and use the 2D version of the MAPYSO code to compute 
a proper radiative transfer integration along each line of sight. 
If we ignore departures from a spherical density distribution, 
this approach would remain strictly exact in the case of solid-body rotation, 
since the velocity difference between 
any couple of points projected on the axis joining these points is insensitive 
to the addition of a solid-body rotation component \citep[cf.][]{Ward01}. 
In practice, however, the rotation observed here departs from solid body 
and the density distribution is not spherical. We therefore assume that, 
to first order, the line excitation is much more sensitive to the 
density distribution (averaged over angles) than to the velocity field. 
(In particular, we have checked that the profiles of excitation temperature 
are essentially insensitive to the infall velocity field as long as the
turbulent velocity dispersion is of the same order as the infall velocity.) 
The rotation velocity field has nevertheless important effects on the shape of 
the line profiles, which we properly take into account here.

%
\subsection{Two regimes of rotation}
\label{analyze_2d}

\begin{figure*}[!ht]
\centering
\includegraphics[width=16.cm,angle=0]{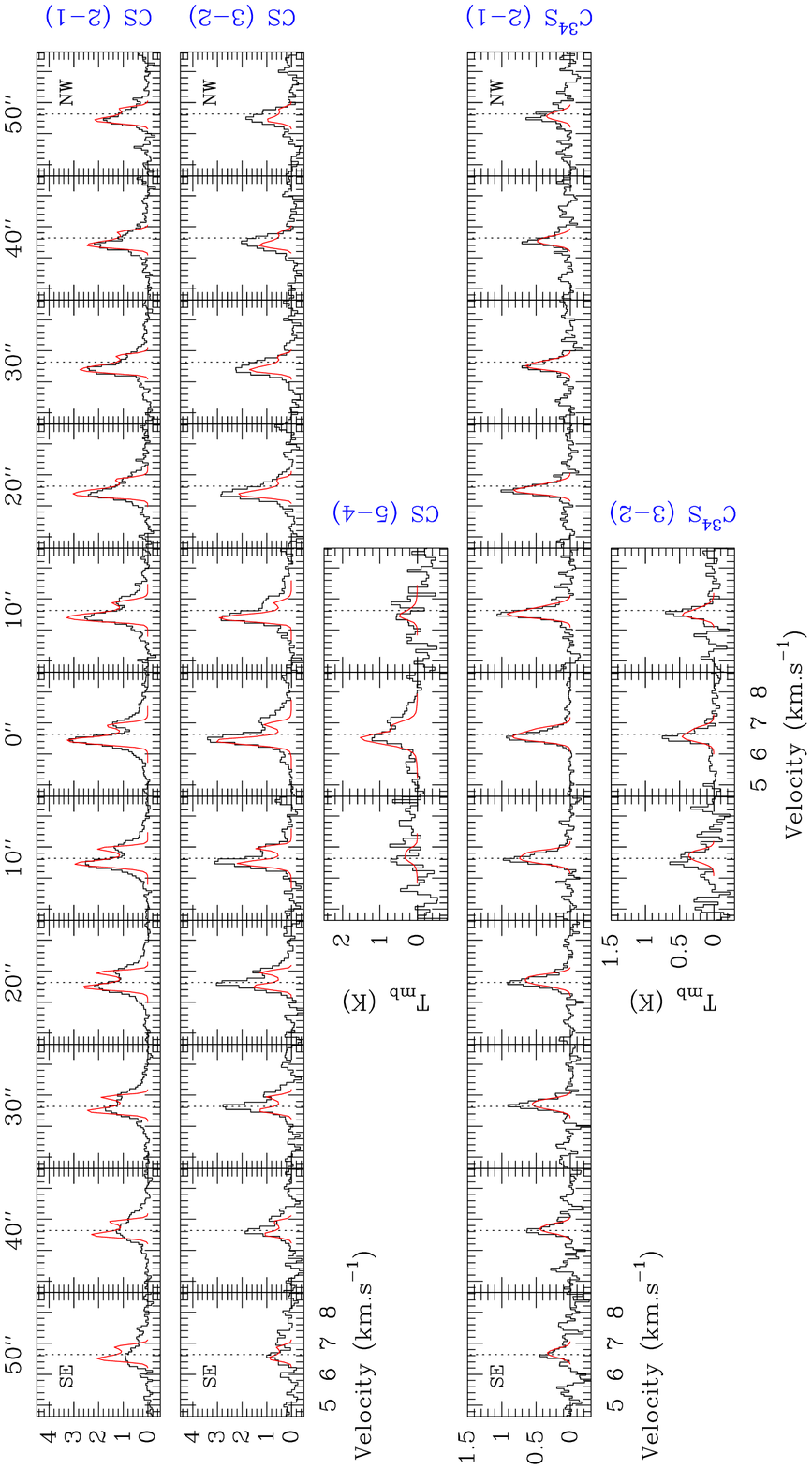}
 \caption{Same as Fig.~\ref{compare_1d} but with synthetic spectra 
corresponding to the `best-fit' 2D model including differential 
rotation described in \S~\ref{analyze_2d} (cf. Fig.~\ref{model_1d}
and Fig.~\ref{model_vel} for model parameters).}
 \label{compare_2d}
\end{figure*}

The cylindrical rotation velocity field that we have added to the 1D 
spherical model of \S~\ref{analyze_1d} is shown in Fig.~\ref{model_vel}c.
The rotation axis is taken to coincide with the outflow axis, at an 
inclination angle $i = 50^{\circ}$ to the line of sight (see 
\S~\ref{intro_iram04191}) and a position angle P.A.~$= 28~\degr$ in 
projection onto the plane of the sky (cf. Fig.~\ref{n2h+flow}).

Adding rotation yields two major improvements in the fits to the CS and 
C$^{34}$S spectra (see Fig.~\ref{pvdiagmod} and Fig.~\ref{compare_2d}). 
First, the velocity gradient of 
the small optical depth C$^{34}$S(2-1) line along the direction perpendicular 
to the outflow is well reproduced, as shown in Fig.~\ref{pvdiagmod}. Both the 
model C$^{34}$S(2-1) position-velocity diagram (Fig.~\ref{pvdiagmod}a) 
and the centroid velocity curve agree well with the observations 
(Fig.~\ref{pvdiagmod}b).
 
Second, the blue asymmetry of the model CS(2-1) and CS(3-2) spectra 
is enhanced toward the north-west and attenuated toward the south-east, as 
seen on the observed spectra (see \S~\ref{obs_infall} and 
Fig.~\ref{compare_2d}). 

In agreement with the discussion of the position-velocity diagrams 
(\S~\ref{obs_rotation} above), the present 2D modeling indicates that the 
envelope can be divided into two regions with distinct rotational
characteristics 
(see the radial profiles derived for the rotational velocity and angular 
velocity in Fig.~\ref{model_vel}c and Fig.~\ref{model_vel}d, respectively). 
First, solid-body rotation is ruled out in the outer $3500 < r < 7000$~AU 
radius envelope, where a good fit to the centroid velocity curve is obtained 
with  $v_\mathrm{rot} \propto r^{- 1.5 \pm 0.5}$, corresponding to an angular
velocity $\Omega \propto  r^{- 2.5 \pm 0.5}$. 
Second, in the inner $r < 3500$~AU radius region, our simulations suggest 
a velocity profile $v_\mathrm{rot} \propto r^{0.1 \pm 0.4}$, i.e., 
$\Omega \propto r^{-0.9 \pm 0.4}$, although the form of the 
position-velocity diagram is significantly influenced by the finite 
resolution of the observations. A rotation velocity 
$v_\mathrm{rot} = 0.20 \pm 0.04$ km~s$^{-1}$, corresponding
to $\Omega  = 12 \pm 3$~km~s$^{-1}$~pc$^{-1}$,
is derived at a radius $r = 3500$ AU after correcting for 
inclination\footnote{The reason why the value of $12$~km~s$^{-1}$~pc$^{-1}$
found here for $\Omega$ is larger than the apparent velocity gradient of 
$9$~km~s$^{-1}$~pc$^{-1}$ given in \S~\ref{obs_rotation} is that the 2D model 
shown in Fig.~\ref{pvdiagmod} properly accounts for the finite 
($\sim 25\arcsec $) resolution of the observations.}.

\section{Discussion: Comparison with collapse models}
\label{discuss}

In this section, we first summarize the predictions of collapse models 
and the main constraints derived from our observations 
(\S ~\ref{discuss_models} and \S ~\ref{discuss_summ}), and then discuss 
the applicability of various models to IRAM~04191 (\S ~\ref{discuss_shu}
to \S ~\ref{discuss_magnet}). Some implications for the distribution and 
evolution of angular momentum during protostellar collapse are discussed  
in \S ~\ref{discuss_ang}.

\subsection{Overview of model predictions}
\label{discuss_models}

\citet{Whitworth85} have shown that there is a two-dimensional
continuum of similarity solutions to the problem of isothermal
spherical collapse.  In this continuum, the well-known solutions
proposed by \citet{Shu77} and \citet{Larson69}-\citet{Penston69} represent two
extreme limits. All isothermal similarity solutions share a universal
evolutionary pattern.  At early times ($t < 0$), a compression wave
(initiated by, e.g., an external disturbance) propagates inward at the
sound speed, $\as$, leaving behind it a $\rho(r) \propto r^{-2}$ density
profile and a uniform infall velocity field. This compression wave has 
zero amplitude in the limiting case of the Shu `inside-out' collapse 
solution.
At $t = 0$, the compression wave reaches the center and a
point mass forms which subsequently grows by accretion.  At later
times ($t > 0$), this wave is reflected into a rarefaction or
expansion wave, propagating outward (also at the sound speed) through
the infalling gas, and leaving behind it free-fall density and velocity  distributions (i.e., $\rho(r) \propto r^{-1.5}$ and $v(r) \propto r^{-0.5}$). 
The various solutions
can be distinguished by the {\it absolute} values of the density and
velocity at $t \sim 0$. 
The \citet{Shu77} solution has the $\rho(r) = (\as^2/2\pi\,G)\ r^{-2}$ 
density distribution of a static ($v = 0$) singular isothermal sphere
(SIS) at $t =0$, while the Larson-Penston (1969) solution is $\sim 4.4$ times
denser and far from equilibrium ($v \approx -3.3\ \as$).  
The recent finding of inward motions of subsonic
amplitude $\sim 0.02-0.10$ km~s$^{-1}$,  extended over $\sim 0.1$~pc 
($\sim 20000$~AU) in the prestellar core L1544 
\citep[see][]{Tafalla98,Williams99,Caselli01a} suggests that true 
protostellar collapse in Taurus proceeds in a manner which is
neither the Shu nor the Larson-Penston flow, and is perhaps more reminiscent
of an intermediate similarity solution.

In practice, the initial conditions for fast protostellar
collapse are not strictly self-similar and involve a density profile
that is flat at small radii \citep[e.g.][]{Ward94,Andre96} and bounded or
sharp-edged at some outer radius $\rout $ \citep[e.g.][]{Motte98, Bacmann00}
like a finite-sized
Bonnor-Ebert isothermal sphere \citep*[e.g.][]{Bonnor56, Alves01}. 
A number of recent
numerical (magneto)hydrodynamic simulations or simplified analytical
calculations attempt to describe the collapse in such a situation,
either in the absence \citep*[e.g.][]{Foster93,Henriksen97,Masunaga98,
Hennebelle02} or in
the presence \citep[e.g.][]{Tomisaka96,Basu97,Safier97,Li98,Ciolek98} of 
magnetic fields.  
The Larson-Penston similarity solution is found to describe the collapse
quite satisfactorily near $t= 0$ (at least for small radii), but the
Shu solution is more adequate at intermediate $t \geq 0$ times, before
the expansion wave reaches the edge of the initial, pre-collapse dense core.

When rotation is included, a rotationally-supported disk develops at the 
center of the infalling envelope during the accretion phase 
(i.e. at $t > 0$). The size scale of this disk is determined  
by the centrifugal radius, $R_C $, which defines the position 
where the centrifugal force balances gravity in the equatorial plane. 
Strong departures from a spherical density distribution and a purely 
radial inflow in the envelope are expected to occur on size scales of order
(or smaller than) $R_C $ \citep[cf.][]{Chevalier83,Hartmann98}.
Most collapse models predict that $R_C \equiv j^2/Gm$ 
should increase with time as material of higher and higher specific angular momentum $j = \Omega \, R^2$ falls in, but the exact dependence 
on time $t$, or alternatively accumulated central mass $m$,   
varies from model to model, according to the distributions of mass and 
angular momentum at $t = 0$. For instance, $R_C $ scales as $m^3$ or $t^3$
in the Shu model (\citeauthor*{Terebey84}), which assumes solid-body rotation 
at point mass formation. By contrast, the dependence of $R_C $ on $m$ is 
only linear in the magnetically-controlled collapse model of \citet{Basu98} 
\citep[see also][]{Krasnopolsky02}.

%
\subsection{Summary of observational constraints}
\label{discuss_summ}

The analysis of our line observations (\S~\ref{simul_1d} and 
\S~\ref{simul_2d}) indicates that 
both the infall and rotation velocity fields of the IRAM~04191 envelope
are characterized by an inner and an outer regime (see Fig.~\ref{model_vel}).\\
Our 1D radiative transfer simulations (\S~\ref{simul_1d}) 
indicate that the infall velocity profile is flat with $v_\mathrm{inf} \sim 
0.1$ km~s$^{-1}$ between $r_\mathrm{i} \sim 2000$ AU and $r_\mathrm{i,o} \sim 
10000-12000$ AU. Higher infall velocities at radii $r < r_\mathrm{i}$ are 
suggested the CS observations, which are consistent with a free-fall velocity
field ($v_\mathrm{inf} \propto r^{-0.5}$) at $r < r_\mathrm{i}$.
The width of the optically thin C$^{34}$S lines strongly constrains
$v_\mathrm{inf}$ to be $\simlt 0.15$ km~s$^{-1}$ at $r \sim r_\mathrm{i}$ 
(cf. Fig.~\ref{model_vel}a).  

The position-velocity diagrams observed in optically thin lines 
(\S~\ref{obs_rotation}) show that the envelope is differentially 
rotating with an angular velocity profile $\Omega \propto r^{-2.5 \pm 0.5}$  
between $r_\mathrm{m} \sim 3500$ AU and $r_\mathrm{m,o} \sim 7000$ AU,
and $\Omega \sim 12$ km~s$^{-1}$~pc$^{-1}$ at $r \sim r_\mathrm{m}$.
Although the limited spatial resolution of our observations prevents us from 
deriving accurate values in the inner region, the rotation profile is 
definitely shallower for $r < r_\mathrm{m}$ (cf. Fig.~\ref{model_vel}d). 
A follow-up interferometric study is underway to provide more 
accurate constraints on the angular velocity in this inner region. 

Although the two critical radii $r_\mathrm{i}$ and $r_\mathrm{m}$ differ 
by less than a factor of $\sim 2$, it is unclear whether they 
are physically related or not.
The inner $r < r_\mathrm{i}$ envelope may correspond to the 
free-fall region developing inside the expansion wavefront, 
while the outer $r > r_\mathrm{i}$ envelope may be dominated by the flat, 
extended inward velocity field set up by the compression wave at $t < 0$
(see \S~\ref{discuss_models} above).
If the current radius of the expansion wave is indeed $\sim r_\mathrm{i}$, 
then the age of the central protostellar object should be 
$t \sim r_\mathrm{i}/a_s \simlt 5 \times 10^4$~yr (assuming a propagation 
speed $a_s \sim 0.2$~km~s$^{-1}$), in rough agreement with the estimated age 
of $\simlt 3 \times 10^4$~yr (\S~\ref{intro_iram04191}).
On the other hand, the fastly rotating $r < r_\mathrm{m}$ region
may be a dynamically collapsing `supercritical' core in the process of 
decoupling from the ambient medium, and the outer $r_\mathrm{m} < r < 
r_\mathrm{m,o}$ envelope may be a transition region between the protostar and 
the background cloud (see \S~\ref{discuss_magnet} below).

%
\subsubsection{Centrifugal support}
\label{discuss_cent}

\begin{figure}
\resizebox{\hsize}{!}{\includegraphics[width=6cm,angle=270]{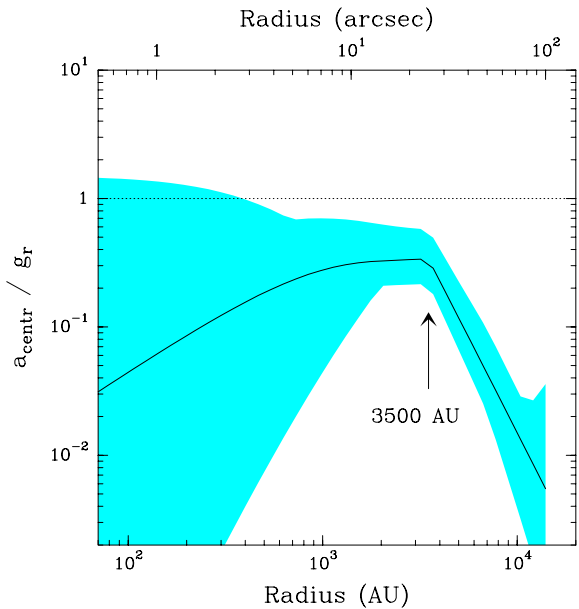}}
\caption{Ratio of centrifugal to gravitational acceleration inferred in the 
IRAM~04191 envelope based on our 2D (\S~\ref{simul_2d}) radiative transfer 
modeling. The dashed area shows the estimated domain where the models match
the CS and C$^{34}$S observations reasonably well. The solid line shows the 
profile of $a_\mathrm{centr}/g_\mathrm{r}$ in the 2D envelope model presented in 
Fig.~\ref{model_vel}, \ref{pvdiagmod} and \ref{compare_2d}, assuming 
$M_\star = 0.05\, M_\odot$.}
\label{ratioacc}
\end{figure}

Using our derived model for the structure and kinematics of the 
IRAM~04191 envelope (e.g. Fig.~\ref{model_1d} and Fig.~\ref{model_vel}), 
we can estimate the dynamical importance of rotation in the envelope.
The ratio of the centrifugal acceleration 
$a_\mathrm{centr} = v_{\mathrm{rot}}^2 / r$ to the local gravitational field 
$g_\mathrm{r} = G \times [M_{\mathrm{env}}(r)+M_\star] / r^2$ is shown in 
Fig.~\ref{ratioacc}. Here, $M_{\mathrm{env}}(r)$ is the envelope mass within 
radius $r$ corresponding to the density profile of Fig.~\ref{model_1d}a, and 
$M_\star$ is the uncertain mass of the central protostellar object 
$\sim 0.03-0.1\, M_\odot$ (cf. \citeauthor*{Andre99}). We estimate a ratio 
$a_\mathrm{centr}/g_\mathrm{r} \sim 0.4 \pm 0.2$
at the radius $r_\mathrm{m} = 3500$~AU of the rapidly rotating inner envelope,
showing that the centrifugal acceleration is a sizeable fraction of the 
gravitational acceleration in this region. 
Comparable values of $a_\mathrm{centr}/g_\mathrm{r} $ 
are nevertheless obtained in some magnetic models of cloud collapse
\citep[see Fig.~4 of][ and \S~\ref{discuss_magnet} below]{Basu94}.

We estimate the centrifugal radius to be 
$R_\mathrm{c} < 400$~AU, assuming the maximum rotation velocity profile
consistent with the observations (cf. Fig.~\ref{model_vel}c) and a stellar mass
$M_\star = 0.03\, M_\odot$. In principle, a large centrifugal radius should 
imply the presence of a large centrifugal disk around the central object.
\citet{Stahler94} have investigated the initial growth of protostellar
disks in the context of the rotating collapse picture of TSC84
and have shown that a 3-component structure should develop inside
$R_\mathrm{c} $: a dense inner accretion disk of radius 
$R_\mathrm{disk }\sim 0.34 R_\mathrm{c} $, 
a ring where material and angular momentum pile up at $R_\mathrm{disk }$, 
and a low-density outer disk where material travels at high velocity between 
$ R_\mathrm{c} $ and $R_\mathrm{disk }$.
If $R_\mathrm{c} \sim 200$~AU, we may thus expect 
$R_\mathrm{disk } \sim 70$~AU. Our 1.3 mm continuum interferometric 
observations set a firm upper limit to the radius of any dense inner 
circumstellar structure around IRAM~04191: $R_\mathrm{disk} < 10$~AU 
(cf. \S~\ref{obs_pdbi}). This suggests that   
the actual centrifugal radius is significantly smaller than 200~AU 
or else that the presence of a protostellar companion at $r \sim 30$~AU 
may have cleared a substantial gap in the inner accretion disk 
\citep[cf.][]{Jensen96}. \citet{Artymowicz94} show  
that, due to tidal disk truncation in binary systems, any individual disk must 
have an outer radius less than half the binary separation, and any
circumbinary disk must have an inner radius more than roughly twice the 
binary separation. The non-detection of a dense $\sim 70$~AU radius disk  
by the PdBI interferometer could thus be accounted for by this effect
if IRAM~04191 were a protobinary of separation $\sim 30$~AU 
(i.e. $\sim 0.2\arcsec $).

%
\subsection{Problems with the inside-out collapse model}
\label{discuss_shu}

The inside-out, isothermal collapse model described by \citet{Shu77}, and its 
2D extension including rotation (\citeauthor*{Terebey84}), has been widely 
used 
to explain the infall spectral signature seen in the envelopes of low-mass 
protostars \citep[e.g.][]{Zhou93,Myers95,Choi95,Zhou96,Hoger00}. 
It accounts relatively well for the densities and accretion rates measured 
in Taurus protostellar envelopes 
\citep[e.g.][; this paper]{Ohashi99,Motte01}. 
This model is, however, inconsistent with the combined  
density and infall velocity profiles measured for IRAM~04191 (cf. 
Fig.~\ref{model_1d}a and Fig.~\ref{model_vel}a). 
An expansion wave radius $\simgt 10000$ AU is indeed required 
to reproduce the blue asymmetry of the central CS(2-1) spectrum in the context 
of the inside-out collapse picture. But such a large infall radius then 
implies high infall velocities at small radii $r \sim 1500$ AU which in turn
yield central C$^{34}$S(2-1) and C$^{34}$S(3-2) spectra that are 2--3 
times too wide compared to the observed linewidths. 
Conversely, a model with an expansion wave radius of only $\sim 2000$ AU 
yields correct C$^{34}$S(2-1) and C$^{34}$S(3-2) linewidths but fails 
to reproduce the strong asymmetry 
observed in CS(2-1) and CS(3-2).
It is clearly because of the absence of significant inward motions at 
$t < 0$ that the inside-out collapse model cannot account
for the infall velocity profile observed here.
A similar inconsistency with the inside-out collapse model has been 
noted previously in prestellar cores such as L1544 by 
\citet{Tafalla98} and \citet{Lee01}.

%
\subsection{Comparison with other thermal models of collapse}
\label{discuss_extend}

With an age $\simlt 3-5 \times 10^4$ yr for the central IRAM~04191 
protostellar object (see \S~\ref{intro_iram04191} and \S~\ref{discuss_summ}) 
and a sound speed $\as \sim 0.19$ km~s$^{-1}$ 
($T_\mathrm{K} = 10$ K), the radius of the expansion wave must be smaller than
the inner radius $r_\mathrm{i} \sim 2000$~AU of the  
extended region where a flat infall velocity field is 
inferred (see Fig.~\ref{model_vel}a). The kinematics and density structure 
of the outer $r > r_\mathrm{i}$ envelope should thus still reflect 
the physical conditions at $t=0$. 
Accordingly, the observation of substantial infall velocities 
between $r_\mathrm{i}$ and $r_\mathrm{i,o} \sim 11000$~AU 
points to collapse models that are more dynamical than the Shu solution 
and involve the propagation of a finite-amplitude compression wave prior 
to the formation of the central object 
\citep[cf. \S~\ref{discuss_models} and][]{Whitworth85}.
On the other hand, the infall velocities derived in the outer 
($r > r_\mathrm{i}$) envelope are only subsonic 
(approximately half the sound speed), and clearly inconsistent with 
the supersonic infall velocities $\sim 3.3\as$ characterizing the 
Larson-Penston isothermal similarity solution at $t = 0$ 
(cf. \S~\ref{discuss_models}).
The infall velocity field derived here (cf. Fig.~\ref{model_vel}a) is 
suggestive of 
a moderately dynamical collapse model, intermediate between the Shu 
similarity solution (zero-amplitude compression wave at $t < 0$) and the 
Larson-Penston solution (strong compression wave at $t < 0$).

Qualitatively at least, such a moderately dynamical infall velocity field 
resembles that achieved during the collapse of a finite-sized, Bonnor-Ebert 
isothermal sphere. For instance, in their numerical simulations
of the collapse of critically stable Bonnor-Ebert spheres (without magnetic
field or rotation), \citet{Foster93} found infall velocities at $t=0$ 
ranging from $3.3\as$ near the origin to 0 at the outer boundary radius. 
Assuming that the IRAM~04191 dense core was initially a marginally stable
Bonnor-Ebert sphere with a center-to-edge density contrast of $\sim 14$, 
the central density at the onset of collapse must have been 
$\sim 5 \times 10^4$ cm$^{-3}$, given the typical outer density 
$\sim 3 \times 10^3$ cm$^{-3}$ observed in prestellar cores 
\citep[e.g.][]{Bacmann00}. 
Adopting a temperature of 10~K, Fig.~1 of \citet{Foster93} then predicts  
an infall velocity at $t=0$ varying from $\as \sim 0.2$ km~s$^{-1}$
at $\sim 6000$ AU to $0.5\as \sim 0.1$ km~s$^{-1}$ at $\sim 15000$ AU.
Even if this represents a somewhat stronger variation of infall velocity 
with radius than derived in the case of IRAM~04191 (Fig.~\ref{model_vel}a), 
the \citeauthor{Foster93} model clearly provides a much better 
fit to the observations than either the Shu or the Larson-Penston 
similarity solution. 
A definite problem with the critical Bonnor-Ebert model, however, is that 
as much as 44\% of the envelope mass is predicted to flow in supersonically 
at $t = 0$ \citep{Foster93}, and progressively more mass at $t > 0$.
This is much larger than the $\sim 1-10\% $ mass fraction derived in 
\S~\ref{analyze_1d} for IRAM~04191. 
We conclude that spherically symmetric collapse models in which thermal 
pressure provides the only force opposing gravity tend to be too dynamical and 
are only marginally consistent with our observational constraints. 
It is likely that the inclusion of rotation in thermal models would 
improve the comparison with observations and may even account for the 
flattened shape of the protostellar envelope seen in Fig.~\ref{n2h+flow} 
(P. Hennebelle, private communication). However, a generic feature of rotating
thermal collapse models is that, due to simultaneous conservation of energy 
and angular momentum, they
tend to predict similar forms for the rotation and infall velocity profiles at 
radii (much) larger than the centrifugal radius \citep[e.g.][]{Saigo98}.
This is at variance with the steeply declining rotation velocity
profile and flat infall velocity profile we observe beyond 3500~AU 
(a radius much larger than the estimated centrifugal radius -- 
see \S ~\ref{discuss_cent} above).
In fact, {\it the strong decline of the rotation velocity profile beyond 
3500~AU  suggests that angular momentum is not conserved in the outer 
envelope.} In the next subsection, we propose that this results from 
magnetic braking.

%
\subsection{Comparison with magnetically-controlled collapse models}
\label{discuss_magnet}

Magnetic ambipolar diffusion models \citep[e.g.][ -- hereafter 
\citeauthor*{Basu94}, \citeauthor*{Basu95a}, 
\citeauthor*{Basu95b}]{Ciolek94,Basu94,Basu95a,Basu95b} are another class of 
models which yield non-zero inward velocities in an extended region 
prior to point mass formation. Ambipolar diffusion has been invoked 
by \citet{Ciolek00} to explain the extended inward motions 
observed in the Taurus prestellar core L1544 
\citep[see][]{Tafalla98,Williams99,Caselli01a}. 
The models start from a magnetically
subcritical cloud, initially supported against gravitational collapse 
by a static magnetic field, and predict an evolution in two phases. 
During the first, quasistatic phase, the subcritical 
cloud contracts along directions perpendicular to the field 
lines through ambipolar diffusion. The gravitationally-induced inward drift 
of the neutral species is slowed down by collisions with the ions which are   
well coupled to the magnetic field. 
The central mass to magnetic flux ratio increases with time, until   
it reaches the critical value for collapse, $(1/2\pi)G^{-0.5}$. 
The inner region of the cloud then becomes magnetically supercritical 
and collapses dynamically, while the outer envelope remains subcritical. 
This effect introduces a spatial scale in the  
collapse process, which corresponds to the boundary between the magnetically 
supercritical inner core and the subcritical outer envelope. 
These two regions are characterized by distinct rotational properties. 
The supercritical inner core evolves with conservation 
of angular momentum and rapidly spins up. It achieves a power-law angular 
velocity profile $\Omega (r) \propto \Sigma (r) \propto 1/r $ at $t=0$, 
where $\Sigma $ is the (mass) column density \citep[e.g.][]{Basu97}.
By contrast, due to magnetic braking, 
the subcritical envelope loses (part of) its angular momentum on the 
timescale $\tau_J \approx t_{ff}^{b} $ (where $t_{ff}^{b} $ 
is the free-fall time at the density of the background -- cf. 
\citeauthor*{Spitzer78} and \citeauthor*{Tomisaka00})
and is progressively brought to near corotation with the background 
medium, assumed to rotate at the uniform rate $\Omega_b $
(e.g. \citeauthor*{Basu94}). This separation generates a break 
in the angular velocity profile at the radius $R_{crit}$ of the magnetically 
supercritical core.

Qualitatively at least, the rotation and infall profiles of the 
IRAM~04191 envelope (see Fig.~\ref{model_vel} and \S~\ref{discuss_summ}) 
can be accounted for in the framework of 
such magnetic models\footnote{Strictly speaking, the ambipolar diffusion 
models of Basu \& Mouschovias describe the evolution of core properties
only during the prestellar phase ($t<0$), while the properties of the 
IRAM~04191 envelope beyond $r_\mathrm{i} \sim 2000$~AU are representative
of the physical conditions at $t \approx 0$ (cf \S ~\ref{discuss_extend}).
However, the models are easily extrapolated 
to $t=0$ by considering the limiting profiles achieved for an infinite central
density \citep[cf.][]{Basu97,Basu98}.}
if we identify $R_{crit}$ with the radius 
$r_\mathrm{m} \sim 3500$~AU beyond which the observed rotation profile exhibits
a marked steepening (cf. Fig.~\ref{model_vel}c).
Indeed, the angular velocity profile of the inner $r < r_\mathrm{m}$ region is 
consistent with the $\Omega \propto 1/r $ power law expected in the 
supercritical core at $t =0$.
Furthermore, the models predict a steepening of the 
rotation profile beyond $R_{crit}$ when magnetic braking does not have  
enough time to bring the system to corotation with the background 
before the formation of the supercritical core. In the parameter study 
presented by \citeauthor*{Basu95a} and \citeauthor*{Basu95b}, this happens 
in models 6 and 8, for two different reasons.
First, if the cloud is already critical (or close to critical)
initially near the center (but still 
subcritical in its outer parts) as in model~6, there is no quasistatic phase 
and the dynamical contraction of the supercritical core starts right away 
on a timescale $\sim t_{ff}^{c} $ (free-fall time at cloud center) 
shorter than $\tau_J $.
This produces a transition region beyond $R_{crit}$ with a steep angular 
velocity profile between the supercritical inner core and the  
background outer region (see Fig.~6b of \citeauthor*{Basu95b}).
At the same time, the collapse of the supercritical core is retarded by the 
magnetic forces so that supersonic infall velocities develop only very close 
to the center (at $r \simlt 0.01 \times R_{crit}$).\\
Second, if the ionization fraction is low enough that the magnetic braking
timescale is only slightly shorter than the ambipolar diffusion timescale 
as in model 8 of \citeauthor*{Basu95a}, then the supercritical core can begin
its dynamical evolution with a rotational angular velocity much larger than that
of the background. This also results in the apparition of a transition region
with a steep rotation profile between the supercritical core and the external  
background (cf. Fig.~5 of \citeauthor*{Basu95a}).
In model~8 of \citeauthor*{Basu95a}, 
the outer radius of the transition region is 
$\sim 3 \times R_{crit} $, which is consistent with the extent of the
$\Omega \propto r^{-2.5}$ zone in Fig.~\ref{model_vel}d.
In this model, the infall velocity becomes supersonic for 
$r \simlt 0.3 \times R_{crit}$ and exhibits a relatively flat profile
beyond $ R_{crit} $, also in agreement with the observational constraints 
of Fig.~\ref{model_vel}a.
The ionization degree $x_i = n_i/n(\mathrm{H}_2)$ assumed in the model 
is low ($2 \times 10^{-9}$ at a density of $10^{6}$~cm$^{-3}$), 
but comparable to the value derived at the center of the 
IRAM~04191 envelope in \S ~\ref{obs_ionization}. 

Quantitatively, however, it is more difficult to obtain a good match of the
observations of IRAM~04191 with published ambipolar diffusion models. 
These models rotate a factor of $\sim 3-10$ more slowly\footnote{Although 
rotating more slowly than the IRAM~04191 envelope, some of the models are 
characterized by high ratios of centrifugal to gravitational
acceleration at the supercritical radius, comparable to the $0.4\pm 0.2$ ratio
observed here (cf. model 1 of \citeauthor*{Basu94} and model 6 of 
\citeauthor*{Basu95b}). Qualitatively, a faster initial 
rotation rate is not expected to change the evolution significantly 
compared to model 6 of \citeauthor*{Basu95b} (S. Basu, private communication).}
than does the 
IRAM~04191 envelope and have supercritical core radii that are a factor 
$\sim 3-30$ bigger than the observed break radius $r_\mathrm{m} \sim 3500$~AU 
(cf. Table 2 of \citeauthor*{Basu95b}).
Physically, the radius of the magnetically supercritical core corresponds 
to the Jeans length for the density $n_{crit}$ (or equivalently surface 
density $\Sigma_{crit}$) at which ``decoupling'' occurs, i.e., the density 
(or surface density) at which gravity overcomes magnetic support. 
In the disk-like geometry of ambipolar diffusion core models, the critical 
Jeans radius is $R_{crit} = \as^2/(2G\Sigma_{crit})$ and one has 
$n_{crit} = \frac{\pi G}{2\mu m_H} \times (\Sigma_{crit}/\as)^2
= \frac{\pi}{8G\mu m_H} \times (\as/R_{crit})^2 $ (see \citeauthor*{Basu95a}).
For $T_{core} = 7$~K and $ R_{crit} = 3500$~AU, this gives 
$N_{crit} \equiv \frac{\Sigma_{crit}}{\mu m_H} \sim 
9.3 \times 10^{21}$~cm$^{-2}$ and 
$n_{crit} \approx 1.4 \times 10^5 $~cm$^{-3}$, the latter being 
remarkably similar to the volume density estimated at $r = 3500$~AU 
in the envelope (as expected in the 
models -- see, e.g., Fig.~6a of \citeauthor*{Basu95b}). 
If the mass--to--flux ratio $M/\Phi = \Sigma/B$ 
is just critical in the supercritical core, then 
the magnetic field strength of the core should be 
$B_{crit} = 2\pi G^{1/2} \Sigma_{crit} \sim 60\, \mu$G at $r \sim 3500$~AU.
The reason why published magnetic models have larger values of $R_{crit}$
is that their critical ``decoupling'' densities and field strengths 
are typically lower than these estimates 
by factors $\sim 10$ and $\sim 3$, respectively.
Direct Zeeman measurements suggest that the magnetic field strength 
is only $\sim 10\, \mu $G in the low-density ($\sim 10^4$~cm$^{-3}$) 
outer parts of prestellar 
cores such as L1544 \citep[][]{Crutcher99,Crutcher00}. Only relatively 
poor upper limits ($< 100\, \mu $G) exist for the field strength in the 
central parts of these cores \citep[][]{Levin01}.
Albeit quite large the above value $B_{crit} \sim 60 \, \mu $G, which refers
to the $\simgt 10^5$~cm$^{-3}$ region, thus remains realistic. 
Furthermore, it should be noted that published 
ambipolar diffusion models include only a static magnetic field and do not 
take turbulent support into account. If a turbulent magnetic field is present,
then a weaker static field may be sufficient to yield a decoupling 
density as high as $n_{crit} \sim 10^5 $~cm$^{-3}$. 
According to this interpretation, 
the observed radius of ``decoupling'' $\sim 3500$~AU would correspond to 
the cutoff wavelength for MHD waves, i.e., 
$\lambda_A \sim 6200\, \rm{AU} \times (\frac{B}{10\mu G}) \times 
(\frac{n_{H_2}}{3 \times 10^3 \rm{cm}^{-3}})^{-1}$ 
\citep[cf.][]{Mouschovias91}, and
the IRAM~04191 dense core would have initially formed through the dissipation
of MHD turbulence \citep[e.g.][]{Nakano98,Myers98} rather than 
ambipolar diffusion.

%
\subsection{Evolution of angular momentum during protostar
formation}
\label{discuss_ang}

\begin{figure}
\resizebox{\hsize}{!}{\includegraphics[width=6cm,angle=270]{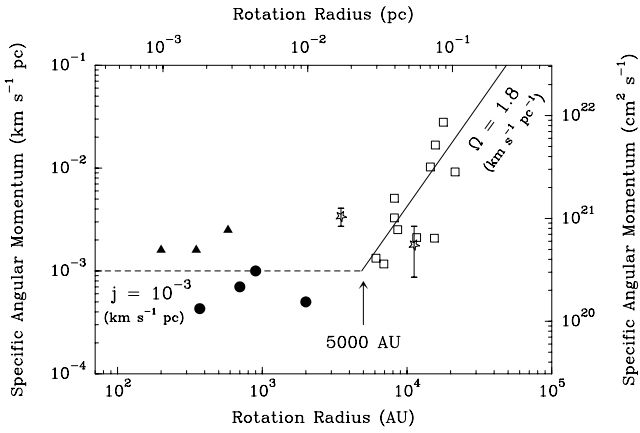}}
\caption{Specific angular momentum as a function of radius for 
the Taurus NH$_3$ dense cores of \citet{Goodman93} (open squares) and the 
rotating ``envelopes'' and ``disks'' observed by Ohashi and collaborators 
around Taurus Class~I sources (filled circles and filled triangles,
respectively). The horizontal dashed line marks the constant specific angular
momentum obtained by \citet{Ohashi99} for his sample of Class~I sources; 
the solid line, corresponding to a constant angular velocity $\Omega = 1.8$ 
km~s$^{-1}$~pc$^{-1}$, shows the best-fit linear correlation determined by us 
for the Taurus NH$_3$ cores of \citet{Goodman93}.
The locations of the fastly rotating inner envelope ($r \sim 3500$ AU) and 
more slowly rotating outer envelope ($r \sim 11000$ AU) in IRAM~04191 are 
marked by open stars, based on the present study. 
\citep[Adapted from][.]{Ohashi99}
}
\label{joverm}
\end{figure}

It is instructive to compare the rotational properties of the 
IRAM~04191 envelope 
with the characteristics of other prestellar and protostellar objects in  
Taurus. Our 2D radiative transfer modeling indicates a rotational velocity  
$v_\mathrm{rot} = 0.20 \pm 0.04$ km~s$^{-1}$ at $r = 3500$ AU, using an 
inclination angle of $i = 50\degr$ (cf. \S~\ref{analyze_2d}).
This corresponds to a local specific angular momentum  
$j = 3.4 \times 10^{-3}$ km~s$^{-1}$~pc at $r_\mathrm{m} = 3500$~AU  
(roughly equal to the mean half-power radius of the 
N$_2$H$^+$ dense core: $FWHM \sim 8800$ AU $\times \, 5200$ AU). 
Quite remarkably, the mean specific angular momentum 
$J/M \sim 1.5 \times 10^{-3}$~km~s$^{-1}$~pc we measure inside 
the rapidly rotating inner $r < r_\mathrm{m}$ envelope of IRAM~04191 
is very similar to the nearly constant $\sim 10^{-3}$~km~s$^{-1}$~pc value 
found by \citet{Ohashi99} for a number of small-scale ``envelopes'' and 
``disks'' around Taurus Class~I sources. More precisely, in the diagram of 
specific angular momentum versus radius presented by \citet{Ohashi97b} 
(see their Fig.~6), IRAM~04191 lies close to the intersection between the 
``dense core'' regime, where the angular velocity is approximately locked 
to a constant background value $\Omega_b \sim 1-2$~km~s$^{-1}$~pc$^{-1}$
(presumably as a result of magnetic braking -- see \citeauthor*{Basu94} and 
\S~\ref{discuss_magnet} above), 
and the ``protostellar'' regime, where the specific angular momentum is 
roughly constant in time\footnote{Assuming that all Taurus sources follow a  
similar time evolution, Fig.~\ref{joverm} may be viewed as an evolutionary 
diagram where the radius plotted on the x-axis represents the ``contraction 
state'' of a core as a function of time. This is conceptually different from 
a plot showing the spatial distribution of angular momentum/velocity in a 
given core at a given time \citep[such as Fig.~1c of][ or Fig.~\ref{model_vel}d
of the present paper]{Basu97}.} 
(cf. Fig.~\ref{joverm}). According to \citet{Ohashi97b}, the 
transition between these two regimes at a radius $\sim 5000$~AU 
(i.e., $\sim 0.03$~pc) characterizes the size
scale for dynamical collapse, inside which evolution proceeds with near 
conservation of angular momentum. 
Interestingly, this size scale is comparable to the radius 
$r_\mathrm{m} \sim 3500$~AU found here for the rapidly rotating 
inner envelope of IRAM~04191. Our suggestion that the inner envelope is  
a magnetically supercritical core decoupling from a subcritical environment
(\S~\ref{discuss_magnet} above) is thus fully consistent with the finding 
and interpretation of \citet{Ohashi97b}.

\section{Summary and conclusions}
\label{concl}

We have carried out a detailed study of the structure and kinematics of the 
envelope surrounding the Class~0 protostar IRAM~04191 in Taurus. Our main 
results and conclusions are as follows:

\begin{enumerate}

\item Extended, subsonic infall motions with 
$v_\mathrm{inf} \sim 0.5\, a_s \sim 0.1$ km~s$^{-1}$, 
responsible for a marked `blue infall asymmetry' in self-absorbed CS and 
H$_2$CO lines, are present in the bulk of the envelope, up to at least 
$r_\mathrm{i,o} \sim 10000-12000$ AU. The observations are also consistent 
with larger infall velocities scaling as $v_\mathrm{inf} \propto r^{-0.5}$
in an inner region of radius $r_\mathrm{i} \approx 2000$~AU.
The corresponding mass infall rate is estimated to be 
$\dot{M}_\mathrm{inf} \sim 2-3 \times a_s^3/G 
\sim 3 \times 10^{-6}$~M$_\odot$~yr$^{-1}$.

\item The protostellar envelope is differentially rotating 
with an angular velocity profile $\Omega \propto r^{-2.5 \pm 0.5}$  
between $r_\mathrm{m} \approx 3500$~AU and $r_\mathrm{m,o} \sim 7000$~AU. 
The rotation profile is shallower, albeit more poorly constrained, 
in the inner $r < r_\mathrm{m}$ region, i.e., 
$\Omega \propto r^{-0.9 \pm 0.4}$. The angular velocity is estimated 
to be $\Omega \sim 12$~km~s$^{-1}$~pc$^{-1}$ at $r \sim 3500$~AU and  
only $\Omega \simlt 0.5- 1$~km~s$^{-1}$~pc$^{-1}$ at $r \sim 11000$~AU. 
The present value of the centrifugal radius is estimated to be 
less than 400 AU.

\item The extended infall velocity profile is inconsistent with the inside-out
collapse picture of \citet{Shu87} and only marginally consistent with 
isothermal collapse models starting from marginally stable equilibrium
Bonnor-Ebert spheres. The latter tend to produce somewhat faster infall
velocities than are observed.

\item The contrast observed between the (steeply declining) rotation velocity 
profile and the (flat) infall velocity profile beyond 
$r_\mathrm{m} \approx 3500$~AU suggests that angular momentum is {\it not} 
conserved in the outer envelope. This is difficult to account for in the 
context of non-magnetic collapse models.

\item Based on a qualitative comparison with magnetic ambipolar diffusion 
models of cloud collapse (e.g. \citeauthor*{Basu94}), we propose that 
the rapidly rotating inner envelope of IRAM~04191 
corresponds to a magnetically supercritical core decoupling 
from an environment still supported by magnetic fields and strongly affected by
magnetic braking. In this view, the outer ($r_\mathrm{m} < r < r_\mathrm{m,o}$)
envelope represents a transition region between the forming protostar and 
the slowly rotating ambient cloud. 
Although published ambipolar diffusion models have difficulty explaining
supercritical cores as small as $R_{crit} \sim 3500$~AU, we speculate that 
more elaborate versions of these models, including the effects of MHD 
turbulence in the outer envelope, would be more satisfactory. 

\item Interestingly, the steepening of $\Omega(r)$ in IRAM~04191 
occurs at a radius comparable to the $\sim 5000$~AU scale inside which 
the specific angular momentum of Taurus dense cores appears to be conserved 
\citep[cf.][ and Fig.~\ref{joverm}]{Ohashi97b}. 
Our results therefore support \citet{Ohashi97b}'s proposal that 
$r \sim 5000$~AU represents the typical size scale for dynamical collapse 
in Taurus. More generally, we suggest that the rotation/infall properties
observed here for IRAM~04191 are representative of the physical conditions
prevailing in isolated protostellar envelopes 
shortly ($\sim 10^4$~yr) after point mass formation.

\end{enumerate}

\begin{acknowledgements} 
We would like to thank Shantanu Basu for enlightening discussions on  
ambipolar diffusion models and Carl A. Gottlieb for providing his 
laboratory measurements of the CS and C$^{34}$S frequencies prior 
to publication. We acknowledge the contribution of Aurore Bacmann during 
the 1999 observing run at the 30m telescope.
We are also grateful to the IRAM astronomers in Grenoble for their help 
with the Plateau de Bure interferometric observations.   

\end{acknowledgements}

\normalsize
\vspace{1cm}
{\bf Appendix: Characteristics of the MAPYSO code}
\vspace{0.2cm}

The numerical code we have used first calculates the non-LTE level populations 
with a 1D (spherical) Monte-Carlo method \citep{Bernes78,Bernes79}. 
Radiative transfer along each line of sight and convolution with the 
antenna beam, approximated by a Gaussian, are then computed with the MAPYSO 
package \citep{Blinder97}. The latter works in both 1D and 2D geometry.

We have tested the Monte Carlo code for two test problems (1 and 2) available 
on the web page of the workshop on Radiative Transfer in Molecular Lines 
held in Leiden in May 1999 
(\mbox{http://www.strw.leidenuniv.nl/$\sim$radtrans/}). 
These tests correspond to a low-abundance and high-abundance HCO$^+$ 
12-level problem, respectively, in the context of the \citet{Shu77}
spherical collapse model. The level populations and 
the excitation temperatures computed by our Monte-Carlo code without any 
reference field \citep[see][]{Bernes79,Pagani98} agree quite well with those 
calculated by the workshop participants. The only significant difference 
occurs for the high-abundance case in the central region 
(inside $\sim$ 1300 AU) where our HCO$^+$(2-1), (3-2), and (4-3) 
excitation temperatures are lower by $\sim 20 \%$ compared to 
the main group results.

The CS and C$^{34}$S Monte-Carlo calculations reported in \S~\ref{simul_1d} 
and \S~\ref{simul_2d} 
used 9 levels and 27 concentric shells. This number of levels should be 
sufficient as the 9$^{\mbox{\tiny th}}$ level is 85~K above the ground level 
while the kinetic temperature in the envelope does not exceed 20~K 
(see \S~\ref{mass_tk}). We used the CS collision rates computed in the 
20-300~K range by \citet{Turner92} and extrapolated these to 5-300~K 
with polynomials \citep[][ and N. Evans, private communication]{Choi95}. 
Each simulation was performed without any reference field 
and resulted from two successive Monte-Carlo runs. The first run 
started from LTE, used packets of 1000 model photons, and computed 100 
iterations, reinitializing the counters after each iteration 
\citep[see][]{Bernes79}. It converged rapidly but still suffered from a high
level of statistical noise. The second run improved the convergence and 
reduced the noise level by starting from the output of the first run, 
computing 40 iterations with packets of 40000 model photons, and 
reinitializing the counters after each group of five iterations.  
We checked that this number of iterations was large enough to reach  
convergence on the populations of the first five levels with an accuracy 
better than a few percents.

\end{document}